\documentclass[aps,nofootinbib,preprintnumbers,twocolumn,superscriptaddress,prd]{revtex4-1}

\usepackage[utf8]{inputenc}

\usepackage{amsmath}
\usepackage{amssymb}
\usepackage{bbold}
\usepackage{slashed}
\usepackage{epsfig}
\usepackage{bbm,bm,slashed}
\usepackage{psfrag}
\usepackage{hyperref}
\usepackage{color}
\usepackage{amsfonts}
\usepackage{dsfont}
\usepackage{siunitx}
\usepackage{graphicx}

\usepackage{calligra}

\usepackage[dvipsnames]{xcolor}

\newcommand{\Eqref}[1]{Eq.~\eqref{#1}}
\newcommand{\on}[1]{\operatorname{#1}}
\newcommand{\Call}[1]{\text{\normalfont\calligra \Large #1}}%
\newcommand{\call}[1]{\text{\normalfont\calligra \large #1}}%

\begin{document}

\title{Critical Wess-Zumino models with four supercharges from the functional renormalization group}


\author{Polina Feldmann}
\affiliation{\mbox{\it Theoretisch-Physikalisches Institut, Abbe Center of 
		Photonics, Friedrich-Schiller-Universit{\"a}t Jena,}
	\mbox{\it D-07743 Jena, Germany}}
\affiliation{\mbox{\it Institut f\"ur Theoretische Physik, Leibniz Universit{\"a}t Hannover, Appelstr. 2, D-30167 Hannover, Germany}}

\author{Andreas Wipf}
\email{wipf@tpi.uni-jena.de}
\affiliation{\mbox{\it Theoretisch-Physikalisches Institut, Abbe Center of 
		Photonics, Friedrich-Schiller-Universit{\"a}t Jena,}
	\mbox{\it D-07743 Jena, Germany}}

\author{Luca Zambelli}
\email{luca.zambelli@uni-jena.de}
\affiliation{\mbox{\it Theoretisch-Physikalisches Institut, Abbe Center of 
		Photonics, Friedrich-Schiller-Universit{\"a}t Jena,}
	\mbox{\it D-07743 Jena, Germany}}



\begin{abstract} 

We analyze the $\mathcal{N}=1$ supersymmetric Wess-Zumino model 
dimensionally reduced to the $\mathcal{N}=2$ supersymmetric model 
in three Euclidean dimensions. As in the original model in four dimensions
and the $\mathcal{N}=(2,2)$ model in two dimensions the 
superpotential is not renormalized. This property puts severe
constraints on the non-trivial fixed-point solutions,
which are studied in detail.
We admit a field-dependent wave function renormalization that
in a geometric language relates to a K\"ahler metric.
The K\"ahler metric is not protected by supersymmetry and
we calculate its explicit form at the fixed point.
In addition we determine the exact quantum dimension of the
chiral superfield and several critical
exponents of interest, including the correction-to-scaling exponent $\omega$, within the functional renormalization group approach.
We compare the results obtained at different levels of
truncation, exploring also a momentum-dependent wave function renormalization.
Finally we briefly describe a tower of multicritical models in continuous dimensions.

\end{abstract}


\maketitle

\section{Introduction}
\label{sec:introduction}

In the challenge of understanding
strongly interacting quantum field theories,
progress has often been associated to 
the special role played by
symmetries. 
Among them, conformal symmetry  and supersymmetry have been of
particular relevance in recent developments in
quantum field theory (QFT) and particle physics.
This line of progress has been a perfect embodiment
of the principle that understanding goes along with simplicity.
Yet, complexity is ubiquitous and the development of general
tools to address it is also an important field of research.
Thus, bringing together powerful symmetries 
and general mathematical methods can be fruitful -- the reliability of the latter can
be tested against the exact constraints imposed by the former.

Remarkably, this kind of analysis is still missing in
several simple arenas offered by
QFT. Such is the $\mathcal{N}=1$  four-dimensional Wess-Zumino 
(WZ)
model~\cite{Wess:1973kz}, and its dimensional reduction
to three dimensions, namely the $\mathcal{N}=2$ model in three dimensions.
While the former has been a seminal example of four-dimensional supersymmetric
field theory and heavily influenced particle phenomenology, 
the latter case and its further reduction to two dimensions
have found surprising applications
to statistical systems, mainly thanks to
the phenomenon of universality.

Two dimensional minimal conformal models with $\mathcal{N}=(2,2)$
supersymmetry~\cite{Fateev:1985mm,DiVecchia:1986fwg,Mussardo:1988av}
have been related to two-dimensional self-dual critical points
of $\mathbb{Z}_N$-symmetric statistical systems, and are benchmark examples
of exactly solvable strongly interacting QFTs.
The relation between these field theories 
and statistical models has even been extended away from 
criticality~\cite{Cecotti:1992vy,Cecotti:1992rm}.
Though these successes deeply rely on the infinite dimensional 
superconformal symmetry
that is special of two dimensions, some of them are based on
another remarkable property of these models, which is present
also in absence of conformal symmetry and in higher dimensions: the
nonrenormalization of the superpotential.
This has been used, for instance, 
to obtain a generic classification of $\mathcal{N}=2$ superconformal models
in two dimensions~\cite{Vafa:1988uu}, through the language of
Landau-Ginzburg effective Lagrangians and their renormalization group (RG) flow.
The latter 
has also been discussed in the framework of conformal perturbation theory in
 Ref.~\cite{Bienkowska:1991uc}.
 
The
nonrenormalization of the superpotential
has been discovered by means of perturbation theory in four dimensions~\cite{Wess:1973kz,Iliopoulos:1974zv,Grisaru:1979wc}.
Later on it has been further analyzed  
nonperturbatively by holomorphy arguments~\cite{Seiberg:1993vc} 
and by algebraic methods~\cite{Flume:1999jf}.
These results refer to the four-dimensional $\mathcal{N}=1$ WZ model,
which is expected to be trivial, i.\,e.~incomplete, beyond pertubation theory.
Yet, similar arguments can be used in less than four dimensions where such a problem is not present.

In fact, dimensional arguments suggest that
non-trivial scale-invariant models with four supercharges
should exist below four dimensions.
Supporting evidence comes from comparison with 
the non-supersymmetric counterpart of this field theory,
the bosonized Nambu$-$Jona-Lasinio model
(the Gross-Neveu model with U$(1)$ chiral symmetry) which can be 
studied perturbatively by means of
$\epsilon$ and $1/N_f$ expansions~\cite{Rosenstein:1990nm,Gracey:1993cq,Gracey:1994ux,Kleinert:1998kj}.
The existence of a continuous phase transition in this model has also been 
confirmed by extensive Monte Carlo studies~\cite{Karkkainen:1993ef,Hands:1992be,Hands:1995jq,Barbour:1999mc,Hands:2001cs,Christofi:2006zt,Chandrasekharan:2013aya,Hesselmann:2016tvh}.
Under the assumption that the critical point survives in
the $N_f\to 1/2$ limit, i.\,e. 	with one Majorana fermion in four dimensions or one Dirac fermion in three dimensions,
the resulting critical field theory 
is expected to enjoy supersymmetry, thus representing
a non-trivial $\mathcal{N}=2$ scale-invariant 
WZ model 
in three dimensions~\cite{Strassler:2003qg}.
The possibility that this supersymmetric fixed point exists in three dimensions
 and the constraints on it stemming from the 
nonrenormalization of the superpotential have already been discussed
through holomorphy in Ref.~\cite{Aharony:1997bx}.

This three-dimensional critical QFT
has received much attention recently, as a candidate for
the emergence of supersymmetry in the long-distance
physics of condensed matter systems~\cite{talkEmergentSusy}.
Concerning such systems, several specific 
proposals were made~\cite{Lee:2006if,Lee:2010fy,Ponte:2012ru,Grover:2013rc}.
For dedicated QFT analyses of 
emergent supersymmetry see also Refs.~\cite{Sonoda:2011qd,Fei:2016sgs,Zerf:2016fti,Gies:2017tod,Zhao:2017bhw}.
These works brought constructive evidence about the
existence of scale-invariant three-dimensional models with
$\mathcal{N}=1$ and $\mathcal{N}=2$ supersymmetry,
but for the $\mathcal{N}=2$ model only through a 
perturbative $\epsilon$ expansion about four dimensions.
Since the non-trivial fixed point is at strong coupling
in three dimensions, nonperturbative techniques are needed to
investigate its properties.
Among them, the conformal bootstrap has 
put bounds on several quantities of interest,
supporting the conjecture that the fixed point exists
between four and two dimensions and
enjoys superconformal symmetry~\cite{Bobev:2015vsa,Bobev:2015jxa}.
In addition, exact determinations of the sphere free energy~\cite{Jafferis:2010un}
and of the coefficient $C_T$ entering the stress-tensor two-point function~\cite{Imamura:2011wg,Closset:2012ru,Nishioka:2013gza} 
have been provided through localization.

Clearly, other studies with different tools would be helpful to
get a more comprehensive picture of WZ models with four supercharges.
This is the goal of the present work,
where we adopt a nonperturbative functional renormalization 
group (FRG) method that can be 
applied to study strongly interacting 
QFTs in a continuous
number of dimensions. 
The FRG has the virtue 
of including both perturbative and nonperturbative approximations
in one analytic differential framework, which strongly resembles Schwinger-Dyson equations.
Since its very birth~\cite{Wilson:1973jj,Wegner:1972ih} this method has been
extensively applied to critical phenomena, especially in three-dimensional Euclidean spacetime,
both for spin-zero and spin-one-half field theories.
Some illustrative examples and a thorough discussion of the method
can be found in several reviews~\cite{Berges:2000ew,Kopietz:2010zz,Gies:2006wv}.

This framework has been shown to
yield compelling results in three-dimensional critical Yukawa models 
at zero temperature and
density~\cite{Rosa:2000ju,Hofling:2002hj,Strack:2009ia,Gies:2009da,Braun:2010tt,Sonoda:2011qd,Scherer:2012nn,Janssen:2012pq,Janssen:2014gea,Gehring:2015vja,Vacca:2015nta,Classen:2015mar,Classen:2017hwp}.
A relevant subclass of such systems --
supersymmetric models -- has also been under the focus of the FRG.
Especially in three dimensions, the $\mathcal{N}=1$ WZ model has been
studied in greater detail~\cite{Synatschke:2010ub,Heilmann:2014iga,Hellwig:2015woa,Gies:2017tod},
but also O(N) models have been 
addressed~\cite{
Heilmann:2012yf}.
Concerning theories with four supercharges, applications have been essentially 
limited to reproducing the nonrenormalization of the superpotential in
four~\cite{Rosten:2008ih,Sonoda:2009df}, three~\cite{Osborn:2011kw} 
and two dimensions, and to an analysis of the two-point 
function in the latter case~\cite{Synatschke:2010jn}.

In contrast, this work is devoted to the construction 
and characterization of non-trivial fixed points.
As such, it is an exploratory study that lays the
basis for a more comprehensive FRG analysis.
In particular, in Sec.~\ref{sec:model section}
we recall the definition of the four and three-dimensional
models and the dimensional reduction that links them.
In Sec.~\ref{sec:FRG section} we adopt the three-dimensional
parametrization to detail the FRG framework.
 In Sec.~\ref{sec:constraints} we
discuss the constraints that
the nonrenormalization of the superpotential imposes on the 
fixed points of the RG in arbitrary dimensions,
including the exact determination of the quantum dimension
of the chiral superfield and several critical exponents,
at an infinite tower of non-trivial scale-invariant models.
In Sections~\ref{sec:LPA$'$ section},~\ref{sec:Field-dependent Z section} 
and~\ref{sec:momentumdependent Z section} we analyze
 the three-dimensional non-trivial critical point,
focusing on the correction-to-scaling exponent $\omega$
which is not constrained by supersymmetry, as a case study
for probing the quality of perturbative and nonperturbative
approximations within the FRG.
Section~\ref{sec:LPA$'$ section} contains the simplest computation,
that only accounts for the running of the wave function renormalization.
Section~\ref{sec:Field-dependent Z section} includes the RG flow of
a field-dependent K\"ahler metric, while Sec.~\ref{sec:momentumdependent Z section},
explores the effect of addressing the momentum dependence of the 
K\"ahler metric.
The remaining infinite number of RG fixed points is addressed in
Sec.~\ref{sec:Multicrit}. Here we perform the minimal analysis
to support their existence in continuous dimensions
between three and two, and to show that they describe genuine 
non-Gaussian models; that is, we perform an $\epsilon$ expansion
around the corresponding upper critical dimensions, within the
truncation of a field-dependent  K\"ahler metric.
We conclude in Sec.~\ref{sec:conclusions} with a summary of
our results and an outlook.
Subsidiary information is provided in the Appendix.

\section{The Wess-Zumino model  in four and three dimensions}
\label{sec:model section}

Our starting point is the WZ model  
\begin{align}\label{eq:WessZumino4D}
\begin{split}
\mathcal{L}_4=&\;\partial_\mu\phi\partial^\mu\!\phi^\dagger+\frac{i}{2}\bar{\psi}\slashed{\partial}\psi+f\!f^\dagger\\&+\left\{\frac{\partial W(\phi)}{\partial\phi}f- \frac{1}{4}\bar{\psi}(1-\Gamma_5)\frac{\partial^2W(\phi)}{\partial\phi^2}\psi+\on{h.\!c.}\right\}
\end{split}
\end{align}
in four-dimensional Minkowski spacetime. $W$ denotes an arbitrary holomorphic superpotential of the form
\begin{equation}
W(\phi)=\sum\limits_{n=1}^{\infty}\frac{C_n}{n}\phi^n\, .
\label{eq:polynomialW}
\end{equation}
The fields $\phi$ and $f$ are complex scalars whereas $\psi$ is a Majorana spinor. 

Dimensional reduction allows to ``downscale'' $\mathcal{L}_4$ 
to three Euclidean dimensions
through compactification of the time direction.
Technically, the time dependence of the fields is abandoned and
their canonical dimensions are adjusted to ensure that a three-dimensional integration over \eqref{eq:WessZumino4D} yields a dimensionless result. The obtained expression can be understood as a Lagrangian density in three dimensions. 

To rewrite it in a familiar form it is useful to pick a particular 
representation of the four and three-dimensional Dirac matrices 
$\Gamma^\mu$ resp.~$\gamma^j$:
\begin{equation}
\Gamma^0=\sigma^2\otimes\mathbb{1},\;\;\Gamma^j=\sigma^3\otimes\gamma^j
\end{equation}
We choose the three-dimensional Dirac matrices proportional to the Pauli 
matrices, $\gamma^j=i\sigma^j$.
The equality
\begin{equation}
[\Gamma^i,\Gamma^j]=\mathbb{1}\otimes[\gamma^i,\gamma^j]
\end{equation}
ensures that 
\begin{equation}
\psi=(1,0)^T\!\otimes\psi_1+(0,1)^T\!\otimes\psi_2
\end{equation}
defines three-dimensional (two component) spinors $\psi_{1}$ and $\psi_{2}$.

Introducing the Dirac spinor
\begin{equation}
\tilde{\psi}=\frac{1}{\sqrt{2}}(\psi_1+\psi_2)
\end{equation}
and abandoning the tilde we end up with the three-dimensional, Euclidean $\mathcal{N}=2$ WZ model 
\begin{align}
\begin{split}
\mathcal{L}_3=&\;\partial_j\phi\partial^j\!\phi^\dagger+i\bar{\psi}\sigma^j\partial_j\psi-f\!f^\dagger\\&-\left\{W'f-\frac{1}{2}W''\psi^T\sigma^2\psi+\on{h.\!c.}\right\}.
\label{eq:WessZumino3D}
\end{split}
\end{align}
The Lagrangian density \eqref{eq:WessZumino3D} 
is, up to a total derivative,
invariant under the supersymmetry transformations
\begin{align}
\begin{split}\label{eq:SusyTrafos}
\delta_\alpha\phi&=\sqrt{2}\alpha^T\!\sigma^2\psi,\\
\delta_\alpha f&=i\sqrt{2}\bar{\alpha}\sigma^j\partial_j\psi,\\
\delta_\alpha\psi&=\sqrt{2}(f\alpha-i\sigma^j\partial_j\phi\sigma^2\!\alpha^*).
\end{split}
\end{align}
The three-dimensional model
\ref{eq:WessZumino3D} can be
constructed 
 from the chiral superfield
\begin{align}
\begin{split} \label{eq:sufield}
\Phi(x,\theta,\bar{\theta})=&\on{e}^{\delta_\theta}\!\phi(x)\\
=&\on{e}^{-i\bar{\theta}\sigma^j\theta\partial_j}(\phi+\sqrt{2}\theta^T\!\sigma^2\psi+\theta^T\!\sigma^2\theta f)
\end{split}
\end{align}
with $\delta_\theta$ defined by \eqref{eq:SusyTrafos}. 
The Lagrangian density \eqref{eq:WessZumino3D} is, up to a surface term, identical to
\begin{align}
\begin{split}
\mathcal{L}_3=&-\frac{1}{4}\int\!\on{d}^2\!\theta\on{d}^2\!\bar{\theta}\;\Phi\Phi^\dagger\\&-\left\{\frac{1}{2i}\int\!\on{d}^2\!\theta\; W(\Phi)+\on{h.\!c.}\right\}.
\label{eq:WessZuminoSF}
\end{split}
\end{align}
The super-covariant derivatives take the form
\begin{align}
\begin{split} \label{eq:covderivatives}
D&=i\sigma^j\theta\partial_j+\partial_{\bar{\theta}},\\
\bar{D}&=-i\bar{\theta}\sigma^j\partial_j-\partial_\theta.
\end{split}
\end{align}
These derivatives allow us to write
the most generic Lagrangian density of the $\mathcal{N}=2$ three-dimensional
WZ model  as \cite{Rosten:2008ih}
\begin{align}
\begin{split}
\mathcal{L}=&-\frac{1}{4}\int\!\on{d}^2\!\theta\on{d}^2\!\bar{\theta}\;\mathcal{K}(D,\bar{D},\Phi,\Phi^\dagger)\\&-\left\{\frac{1}{2i}\int\!\on{d}^2\!\theta\; W+\on{h.\!c.}\right\}\,,
\label{eq:fullLagrangeSufield}
\end{split}
\end{align}
where 
$\mathcal{K}$ is an arbitrary real, scalar, analytic function of $\Phi$, $\Phi^\dagger$, and of the covariant derivatives, which act on the fields. 
Though $\mathcal{K}$ is sometimes called the Kähler potential,
we reserve this designation for $K(\Phi,\Phi^\dagger)$, containing only the $D$- and $\bar{D}$-independent contributions to the generalized Kähler potential $\mathcal{K}$. Throughout this paper we stick to real coupling constants $C_n$
in the expansion (\ref{eq:polynomialW}) of the superpotential,
though this is not required by symmetry.
Our conventions are summarized in App.~\ref{sec:appconventions}.

\section{The functional renormalization group}
\label{sec:FRG section}

The modern implementation of the FRG is formulated in terms of
one-particle-irreducible (1PI) correlation functions~\cite{Wetterich:1992yh,Morris:1993qb,Ellwanger:1993mw,Bonini:1992vh}.
This can be adjusted to manifestly preserve supersymmetry,
by taking advantage of the linearization of 
supersymmetry transformations in the off-shell
formulation, involving auxiliary fields $f$,
or in other words by formulating
the Wilsonian cutoff in superspace,
as detailed in Refs.~\cite{Bonini:1998ec,Synatschke:2008pv,Rosten:2008ih}.
In the present work we apply this framework to the
$\mathcal{N}=2$ three-dimensional WZ model,
using the fields and Lagrangians described in the previous section.

Though Eqs.~(\ref{eq:WessZumino3D}) and (\ref{eq:WessZuminoSF})
have been obtained by dimensional torus-reduction
to $d=3$ Euclidean spacetime dimensions,
from here on they will be used for generic $d$.
This amounts to  analytically continuing the
one-loop momentum integrals in the
beta functionals of the model, 
while keeping fixed the parametrization of the dynamics
as encoded in the effective action.
Whether this is compatible with the change of 
parametrization of degrees of freedom through dimensional reduction,
that is whether the dimensional reduction of the effective action
and the analytic continuation of the corresponding beta functionals commute,
will be discussed in Sec.~\ref{sec:momentumdependent Z section}.

Let us introduce the field component vector
\begin{equation}
\Psi(q)=\left(\phi(q),\phi^*(-q),f(q),f^*(-q),\psi^T(q),\psi^\dagger(-q)\right)^T
\end{equation}
in $d$-dimensional momentum space. Fourier conventions are given 
in App.~\ref{sec:appconventions}.
The 1PI formulation of the FRG focuses on the so-called effective average action $\Gamma_k$, a functional interpolating between the action $S$ (for $k\rightarrow\infty$) and the effective action $\Gamma$ (for $k\rightarrow 0$). The flow of 
the scale-dependent average effective action
with the momentum scale $t=\on{ln}(k/k_0)$ is provided by the equation
\begin{equation}
\partial_t\Gamma_k=\frac{1}{2}\on{STr}\left(\partial_t R_k(\Gamma_k^{(2)}+R_k)^{-1}\right)\,,
\label{eq:Wetterich}
\end{equation}
where
\begin{equation}
\Gamma_k^{(2)}(p,q)=\frac{\overset{\rightarrow}{\delta}}{\delta \Psi^\dagger(p)}\Gamma_k\frac{\overset{\leftarrow}{\delta}}{\delta \vphantom{\Psi^\dagger}\Psi(q)}
\end{equation}
and
$\on{STr}$ denotes the supertrace in both spin and momentum labels. The regulator matrix $R_k(q)$ defines a term playing the role of an infrared masslike regularization in the derivation of~\Eqref{eq:Wetterich}:
\begin{equation}
\Delta S_k=\frac{1}{2}\int\!\on{d}^d\!\!q\;\Psi^\dagger(q)\;R_k(q)\;\Psi(q)\, .
\end{equation}
For the present WZ model  the bare action, or the Wilsonian effective action,
enjoys invariance under the supersymmetry transformations of~\Eqref{eq:SusyTrafos}.
When also $\Delta S_k$ respects supersymmetry, this translates into the same symmetry of the average effective action $\Gamma_k$. 

A supersymmetric regulator which is quadratic in the fields
is always of the form
\begin{align}
\begin{split}
\Delta S_k=&-\frac{1}{4}\int\!\on{d}^d\!x\on{d}^2\!\theta\on{d}^2\!\bar{\theta}\;\Phi^\dagger\rho_2(D,\bar{D})\Phi\\
&-\left\{\frac{1}{4i}\int\!\on{d}^d\!x\on{d}^2\!\theta\; \Phi\rho_1(D,\bar{D})\Phi+\on{h.\!c.}\right\}
\label{eq:DeltaS1} 
\end{split}
\end{align}
where the $\rho_i$ are scalar, $t$-dependent 
functions of the covariant derivatives, and $\rho_2$ is Hermitian. 
It can be shown \cite{PolinaFeldmann:2016bfl,Synatschke:2009nm}
that any such $\Delta S_k$ can be simplified to
\begin{align}
\begin{split}
\Delta S_k=&-\frac{1}{4}\int\!\on{d}^d\!x\on{d}^2\!\theta\on{d}^2\!\bar{\theta}\;\Phi^\dagger r_2(-\partial_x^2)\Phi\\
&-\left\{\frac{1}{4i}\int\!\on{d}^d\!x\on{d}^2\!\theta\; \Phi r_1(-\partial_x^2)\Phi+\on{h.\!c.}\right\}
\label{eq:DeltaS2} 
\end{split}
\end{align}
with $t$-dependent regulator functions $r_1$ and
$r_2$, both analytic in $(-\partial_x^2)$,
$r_2$ being additionally real. 
The proof is similar to the one given in \cite{Synatschke:2008pv}.
Choosing also $r_1$ to be real we obtain the block diagonal regulator matrix as composed of the first, bosonic block
\begin{equation}
R_B(q)=
\begin{pmatrix}
q^2r_2(q^2)\mathbb{1} & -r_1(q^2)\sigma^1\\
-r_1(q^2)\sigma^1 & -r_2(q^2)\mathbb{1}
\end{pmatrix}
\label{eq:reg1}
\end{equation}
and the second, fermionic one
\begin{equation}
R_F(q)=
\begin{pmatrix}
r_2(q^2)\sigma^jq_j & r_1(q^2)\sigma^2\\
r_1(q^2)\sigma^2 & r_2(q^2)\sigma^{jT}q_j
\end{pmatrix}.
\label{eq:reg2}
\end{equation}

Imposing supersymmetry
allows us to write the average effective action as
\begin{align}
\begin{split}
	\Gamma_k = &-\frac{Z_{0k}}{4}\int\!\on{d}^d\!x\on{d}^2\!\theta\on{d}^2\!\bar{\theta}\;\mathcal{K}_k(D,\bar{D},\Phi,\Phi^\dagger)\\
	&-\left\{\frac{1}{2i}\int\!\on{d}^d\!x\on{d}^2\!\theta\; W_k+\on{h.\!c.}\right\}
	\label{eq:genericGamma}
\end{split}
\end{align}
with normalization $\partial_{\Phi^\dagger}\partial_{\Phi}\mathcal{K}_k(0)=1$ and 
real $Z_{0k}$ and $\mathcal{K}_k$, compare with~\Eqref{eq:fullLagrangeSufield}. 
The $k$-subscripts of $Z_0$, $\mathcal{K}$ and $W$ indicate a scale-dependence of 
these
quantities. From now on they will be dropped, since we will be concerned with running coupling constants only. Their infinite number renders 
\Eqref{eq:Wetterich} equivalent to an infinite system of differential 
equations. To make practical use of them, the system of equations is usually truncated: starting from a simplified, still supersymmetric ansatz for $\Gamma$, 
\Eqref{eq:Wetterich} is solved up to the order of the ansatz. 

The various truncations employed to obtain the results presented in this paper are introduced in the following sections. 
However, let us anticipate that, projecting onto $\psi=\bar{\psi}=0$ and constant $\phi$ and $f$, and computing the $f$-derivative of the truncated FRG equations at $f=0$, we always find
\begin{equation}
	\partial_t W = 0.
	\label{eq:nonrenormalization}
\end{equation}
To the order of our truncations the superpotential is scale-invariant.
The implications of this nonrenormalization theorem on the
landscape of critical WZ models in various
dimensions are discussed in the next section.

\section{Constraints on superconformal Wess-Zumino models}
\label{sec:constraints}

As recalled in Sec.~\ref{sec:introduction},
the nonrenormalization theorem can be used to constrain
key properties of the putative superconformal $\mathcal{N}=2$ WZ model 
in three dimensions, such as the dimension of the superconformal
chiral primary $\Phi$, that must be equal to its R-charge.
The same must apply in $d=2$, as well as to other superconformal
theories that could exist below the corresponding fractional 
upper critical dimensions in $2<d<3$.
Such constraints straightforwardly 
descend from \Eqref{eq:nonrenormalization}: The exact nonrenormalization of the 
bare dimensionful superpotential translates into a very simple
and exact flow for the dimensionless renormalized one.
The fixed points of these RG equations correspond to scale-invariant theories.

In formulas, we introduce the dimensionless, renormalized fields
\begin{align}
\begin{split}
X&=Z_0^{1/2}k^{(2-d)/2}\Phi\, ,\\
\chi&=Z_0^{1/2}k^{(2-d)/2}\phi\, .
\end{split}
\end{align}
We further specify the average effective action of \Eqref{eq:genericGamma}
such that $\mathcal{K}_k(D,\bar{D},\Phi,\Phi^\dagger)=\mathcal{K}(-\partial_x^2,\Phi,\Phi^\dagger)$.
All our truncations are thus characterized
by the generalized Kähler metric
\begin{align}
\begin{split}
\zeta(-\partial_x^2,\Phi,\Phi^\dagger)=
\partial_{\Phi^\dagger}\partial_\Phi\mathcal{K}(-\partial_x^2,\Phi,\Phi^\dagger)\, .
\end{split}
\end{align} 
The dimensionless and renormalized, and therefore $Z_0$-independent, 
formulation of our ansatzes for~$\Gamma_k$ can thus be expressed in terms
of the dimensionless, renormalized superpotential and generalized Kähler metric
\begin{align}
\begin{split}
w (X)&=k^{1-d}W(\Phi)\, ,\\
\tilde{\zeta}(-\partial_x^2/k^2,X,X^\dagger)&=
\zeta(-\partial_x^2,\Phi,\Phi^\dagger)\, ,
\label{eq:dimlrenorm}
\end{split}
\end{align}
with $\zeta(0,0,0)=1$. 
This rescaling entails corresponding redefinitions of couplings, such that
\Eqref{eq:polynomialW}
becomes
\begin{equation}
w (\chi)=\sum\limits_{n=0}^{\infty}\frac{c_n}{n}\chi^n\, .
\label{eq:polynomialw}
\end{equation}
For consistency, we also introduce dimensionless, renormalized regulator functions
\begin{align}
\begin{split}
\tilde{r}_1(q^2/k^2)&=\frac{1}{kZ_0}r_1(q^2)\, ,\\
\tilde{r}_2(q^2/k^2)&=\frac{1}{Z_0}r_2(q^2)\, .
\label{eq:rescaling_regulators}
\end{split}
\end{align}
In the following, the tildes are omitted.
	
In shifting our attention
to dimensionless interactions,
the anomalous dimension of the fields
\begin{align}
\eta = -\partial_t\ln Z_0\, ,
\label{eq:def_eta}
\end{align}
enters in
the RG equations, which is eventually determined by the fixed-point RG equations.
From \Eqref{eq:nonrenormalization},
the flow of the dimensionless superpotential (\ref{eq:dimlrenorm}) 
results in
\begin{align}
\begin{split}
\partial_tw =(1-d)w +\Delta\, \chi w '\, ,
\end{split}
\label{eq:renormalizedfloww}
\end{align}
where 
\begin{equation}
\Delta=\frac{d-2+\eta}{2}
\end{equation}
denotes the quantum dimension of $\phi$.
At a fixed point, one has to require $\partial_t w_* = 0$ which, 
for non-vanishing $w_*$, is solved by 
\begin{align}
\begin{split}
w_*(\chi)&=\left(c_{\mathfrak{n}*}\chi^\mathfrak{n}\right)/\mathfrak{n}\, ,\\
\eta_{*}&=\frac{2(d-1)-\mathfrak{n}(d-2)}{\mathfrak{n}}\, ,\\
\Delta_*&=(d-1)/\mathfrak{n}\, ,
\label{eq:etas}
\end{split}
\end{align}
with $\mathfrak{n}>0$.
Requiring that all the on-shell 
effective vertices be finite at zero momenta
selects $\mathfrak{n}\in\mathbb{N}$.
In other words, the set of possible
anomalous dimensions at non-Gaussian fixed points 
is quantized by the analyticity of the 
superpotential.~\footnote{Fixed point potentials that
are not smooth at the origin have been discussed
in the context of three-dimensional O(N) models at 
large-N, with~\cite{Bardeen:1984dx,Litim:2011bf,Heilmann:2012yf} and without~\cite{Bardeen:1983rv,Comellas:1997tf,Marchais:2017jqc} supersymmetry,
where this singularity has been interpreted as
the signal of spontaneous breaking of scale
invariance.
They also appear in the UV asymptotics of four-dimensional non-Abelian Higgs 
models~\cite{Gies:2015lia,Gies:2016kkk},
where the singular behavior originates from a Coleman-Weinberg
mechanism.}
For $\mathfrak{n}=1$ the Lagrangian is symmetric under constant shifts of the fields,
and $\eta_*=d$. 
For $\mathfrak{n}=2$ the superpotential $w_*$ contains only a mass term, and
$\eta_*=1$. Let us stress that in these two cases the superpotential 
is noninteracting but the corresponding  $\mathcal{K}_*$
might be non-trivial.
For $\mathfrak{n}\geq 3$ the anomalous dimension  $\eta_*$ is positive
below the respective upper critical dimensions
\begin{equation}
d_\mathfrak{n}=2\,\frac{\mathfrak{n}-1}{\mathfrak{n}-2}\, .
\label{eq:upper_critical_d}
\end{equation}
To decide whether the fixed points are Gaussian or not,
it is necessary to compute other universal quantities such
as the critical exponents.

Linearizing the flow equation \eqref{eq:renormalizedfloww} about a fixed point, i.\,e.~setting $w =w_*+\delta w $ and $\eta = \eta_* + \delta\eta$, gives 
\begin{align}
\begin{split}
\partial_t\delta w &
=(1-d)\delta w +\Delta_*\,\chi\delta w' +
\frac{\delta\eta}{2}\chi w_*'\, .
\label{eq:linwflow}
\end{split}
\end{align}
The critical exponents $\lambda$ arise as eigenvalues of this linear RG operator,
\begin{equation}
\partial_t\delta w_\lambda=\lambda\delta w_\lambda\, .
\label{eq:eigenvalueproblem}
\end{equation}
Since the fixed-point superpotential is a monomial,
an infinite subset of $\lambda$'s can be computed exactly,
even without knowing $c_{\mathfrak{n}*}$,
and the corresponding eigenfunctions are simple powers:
\begin{align}
\begin{split}
\delta w_n&=(\delta c_n \chi^n)/n\, , \quad\quad \forall n\neq \mathfrak{n}\, ,\\
\lambda_n&=1-d+ n\, \Delta_*=(n-\mathfrak{n})\Delta_*\, .
\label{eq:linsoldw}
\end{split}
\end{align}
The analyticity requirement $n \in\mathbb{N}$ quantizes $\lambda$.
Thus \Eqref{eq:linsoldw}
follows from a polynomial ansatz as in \Eqref{eq:polynomialw},
and from the diagonalization of the stability matrix B defined as
\begin{equation}
\partial_t \delta c_n=\sum_{m}B_{nm}\delta c_m\, .
\end{equation}
In other words, \Eqref{eq:linsoldw} holds regardless of $\delta\eta$ thanks
to the existence of the orthonormal basis of monomial functions.
Incidentally, fixed points with noninteger $\mathfrak{n}$
would require a different basis of eigenfunctions,
to provide directions 
orthogonal to the fixed point superpotential and a discrete spectrum analogous to \Eqref{eq:linsoldw}.
The case $n=\mathfrak{n}$ has to be excluded in~\Eqref{eq:linsoldw}
 because it requires the knowledge 
of $\delta\eta$,
which in turn involves the flow of the generalized K\"ahler
potential. 
If $\delta\eta=0$, $n=\mathfrak{n}$ can be included in~\Eqref{eq:linsoldw},
yielding $\lambda_\mathfrak{n}=0$, which corresponds to a marginal 
direction.
The $n=\mathfrak{n}$ case is the most interesting one, since all
other eigenvalues are Gaussian,
in the sense that the level splitting is equal to the dimension of the field.

Thus, a study of the unprotected derivative sector
of the average effective action is necessary for two complementary
reasons: to collect evidence in favor of the existence (i.\,e.~mathematical consistency)
of fixed points fulfilling the constraints of~\Eqref{eq:etas},
and to decide whether these correspond to genuinely non-trivial
superconformal theories, and not simple Gaussian models
to which we assigned the wrong engeneering (classical) dimensions,
as one might conjecture on the basis of~\Eqref{eq:linsoldw}.
This will be the goal of the following sections.

\section{The wave function renormalization}
\label{sec:LPA$'$ section}

Our first truncation of $\Gamma_k$ reads
\begin{align}
\begin{split}
\Gamma_k=&-\frac{Z_0}{4}\int\!\on{d}^d\!x\on{d}^2\!\theta\on{d}^2\!\bar{\theta}\;\Phi\Phi^\dagger\\
&-\left\{\frac{1}{2i}\int\!\on{d}^d\!x\on{d}^2\!\theta\; W(\Phi)+\on{h.\!c.}\right\}.
\label{eq:truncLPA}
\end{split}
\end{align}
This approximation, including a generic superpotential 
and a wave function renormalization 
which is independent of fields as well as momenta,
is often called LPA$'$, since it is a minimal improvement of the local potential approximation (LPA).
The computation of the RG equations is detailed in App.~\ref{sec:appLPA}.
Let us adopt the abbreviation
\begin{equation}
\int_q =\frac{\Omega_d}{(2\pi)^d}\int\limits_{0}^{1}\!\on{d}\!q\;q^{d-1}\
\label{eq:int_q}
\end{equation}
where  $\Omega_d=2\pi^{d/2}/\Gamma(d/2)$ is the surface of a unit $(d-1)$-sphere. 
Then the flow of the wave function renormalization, in terms of  dimensionless renormalized quantities, results in 
\begin{align}
\begin{split}
\eta=4g^2\int_q\frac{h}{v^3}
&\left( \vphantom{\frac{1}{2}} 2hM(\partial_t-q\partial_q-\eta +1)r_1\right.\\
&\quad\left.\vphantom{\frac{1}{2}}-u(\partial_t-q\partial_q-\eta )r_2\right),
\end{split}
\label{eq:flowlpa}
\end{align}
where $h$, $M$, $u$, $v$ and $r_{1,2}$
are functions of $q^2$, and we have used the notations of Ref.~\cite{Synatschke:2010jn},
that is:
\begin{align}
m &= c_2=w ''(0)\,, & g &= c_3=w '''(0)/2\,,\nonumber\\
h&=1+r_2\,, & M&=m+r_1\,,\label{eq:lpa_notations_uv}\\
u&=M^2-q^2h^2\,, & v&=M^2+q^2h^2\,\nonumber
\end{align}
Inserting this result into the RG equation~\eqref{eq:renormalizedfloww}
for the dimensionless renormalized superpotential determines
the beta function of the last missing coupling, $c_\mathfrak{n}$, of the LPA$'$ truncation.

Equation
\eqref{eq:flowlpa} shows how, for $\eta_*\neq 0$, the LPA$'$ approximation
can be appropriate only for the $\mathfrak{n}=3$ fixed point of~\Eqref{eq:etas}.
In fact, since $\eta$ is proportional to $g^2$, 
\Eqref{eq:flowlpa} would predict $\eta_*=0$ for all other values of $\mathfrak{n}$.
For the $\mathfrak{n}=3$ case it consistently accommodates the
$\eta_*=(4-d)/3$ solution of~\Eqref{eq:etas},
and it further provides a description of this model away from criticality.
The simplest piece of information contained in~\Eqref{eq:truncLPA}
	is the first order of the
expansion in $\epsilon=4-d$ around the Gaussian fixed point in four 
dimensions, which is regulator-independent and reads
\begin{align}
\begin{split}
	&\partial_t g=-\frac{\epsilon}{2}g+\frac{3}{8\pi^2}g^3\\
	\Rightarrow\; &g_*^2=\frac{4\pi^2}{3}\epsilon,\;\;\omega=\left.\frac{\partial(\partial_t g)}{\partial g}\right|_* = \epsilon.
\label{eq:epsexpansion}
\end{split}
\end{align}
As a benchmark to which our results will be compared, let us recall
that the four-loop approximation gives~\cite{Zerf:2017zqi} 
\begin{align}
\begin{split}
\omega =&\, \epsilon-\frac{\epsilon^2}{3}+\left(\frac{1}{18}+\frac{2\zeta(3)}{3}\right)\epsilon^3\, \\
&-\frac{1}{540}\left(35-3\pi^4+420\zeta(3)+1200\zeta(5)\right)\epsilon^4.
\end{split}
\label{eq:epsmore}
\end{align}
In a mass-dependent scheme, such as the ones we will discuss, $\partial_t g$ depends not 
only on $g$, but also on $m$ and the perturbatively nonrenormalizable couplings
of the Kähler potential.
Then, a natural generalization of
formula~\eqref{eq:epsexpansion} for the first correction-to-scaling exponent $\omega$ is to proceed to an eigenvalue of the smallest diagonal block of the stability matrix containing $\partial_g(\partial_t g)|_*$. We identify $\omega$ with the smallest positive eigenvalue of this block. 

For later purposes, it is instructive to describe how $\omega$ in
\Eqref{eq:epsexpansion} stems from the FRG equations.
Expansion of \Eqref{eq:flowlpa} to first order 
in $\epsilon$ and $g^2\sim O(\epsilon)$ produces
\begin{equation}
\eta=\frac{g^2}{4\pi^2}\, ,\label{eq:oneloopeta}
\end{equation}
which fixes $\delta \eta$ in terms of $\delta w'''(0)$.
Inserting this into~(\ref{eq:linwflow}) and requiring (\ref{eq:eigenvalueproblem})
leads to the eigenperturbation
\begin{equation}
\delta w_\lambda(\chi)=\delta c_\lambda \chi^{3+\lambda(1+\epsilon/3)}
+\delta w_\lambda'''(0)\frac{\epsilon}{6\lambda}\chi^3\, ,
\label{eq:deltaw_LPA}
\end{equation}
containing three apparently free parameters: $\lambda$, $\delta c_\lambda$ and $\delta w_\lambda'''(0)$.
The latter can be traded in for $\delta \eta_\lambda$. A vanishing
$\delta w_\lambda'''(0)$ corresponds to the quantized solutions
of~\Eqref{eq:linsoldw}.
If instead $\delta w_\lambda'''(0)\neq 0$, then necessarily $\delta c_\lambda=0$ 
and $\lambda=\epsilon$, to ensure that the third derivative of the 
left-hand side of~\Eqref{eq:deltaw_LPA} at the origin is finite, non-vanishing and equal to 
the $\delta w_\lambda'''(0)$ on the right-hand side, which then plays the role of an arbitrary normalization
factor. Thus, we recover the expected result that $\omega$ corresponds to
$\delta w\propto\chi^3$.

The approximation of~\Eqref{eq:truncLPA} includes not only the first-order quantum corrections
in the $\epsilon$ expansion, but also a resummation of some higher order perturbative contributions,
and is applicable in any dimension, thought the quality of its predictions will of course
strongly depend on $d$.
In the present section we apply this ansatz and the corresponding RG equation~\eqref{eq:flowlpa}
to the $d=3$ fixed point that is expected to be continuously connected 
to the $d=4-\epsilon$ solution of~\Eqref{eq:epsexpansion}.
Our RG equations are scheme-dependent, which means that they 
depend on the choice of the regulator functions $r_{1,2}$.
Though universal quantities such as the critical exponents must be scheme-independent, 
truncation of the exact flow equation~\eqref{eq:Wetterich} 
introduces spurious effects which are well known and
long studied in the literature. The most effective way to deal with these issues
is to vary the regulator and to  optimize it
for each different model and approximation~\cite{Liao:1999sh,Litim:2000ci,Litim:2001fd,Litim:2001up,Canet:2002gs}.
For this reason, we are now turning to the computation of critical exponents
in LPA$'$ with two different regulators.
More general approximations and regularizations will be discussed in
Sec.~\ref{sec:Field-dependent Z section} and Sec.~\ref{sec:momentumdependent Z section}.

The dimensionless, renormalized Callan-Symanzik regulator \cite{synatschkeQM} consists of
\begin{equation}
	r_1=1,\;\;r_2=0.
	\label{eq:CallanSymanzik}
\end{equation}
For $m>-1$ and $0<d<4$ equation \eqref{eq:flowlpa} evaluates to 
\begin{equation}
\eta=\left[1-\frac{4(2\pi)^{d-1}\sin(d\pi/2)(m+1)^{5-d}}{g^2\Omega_d(4-d)(d-2)}\right]^{-1} .
\end{equation}
Together with the flow equation~\eqref{eq:renormalizedfloww} of the superpotential,
this provides a fixed point at $\eta_*=(4-d)/3$ and
\begin{equation}
g_*^2=-\frac{4(2\pi)^{d-1}\sin(\frac{d\pi}{2})}{\Omega_d(d-1)(d-2)}\, ,
\end{equation}
with only one non-Gaussian critical exponent,
\begin{equation}
\omega = \frac{(4-d)(d-1)}{3}=\epsilon-\frac{\epsilon^2}{3}\, ,
\end{equation}
in agreement with the second order of the $\epsilon$ expansion.
This agreement, though remarkable, appears to be accidental.
In three dimensions this amounts to $\omega=2/3$.

A widely used regulator class, which we call of Litim type,
was shown to fulfill an optimization criterion
for several fermionic systems \cite{Litim:2001up}.
We choose such regulators mainly for computational 
convenience.
Since Litim-type regulator functions have a nonanalytic 
momentum dependence, they violate the assumptions under which we
argued that \Eqref{eq:DeltaS1} provides the most generic supersymmetric regularization, see Sec.~\ref{sec:FRG section}.
Yet, using this scheme, in the frame of the employed truncations we encounter no ensuing anomalies. 

The dimensionless, renormalized Litim-type regulator I has the form
\begin{equation}
r_1=0,\;\;r_2=\left(\frac{1}{q}-1\right)\Theta(1-q^2).
\label{eq:litimreg}
\end{equation} 
Evaluating \eqref{eq:flowlpa} in $0<d<4$
gives
\begin{equation}
\eta=(d-1)\left[1-\frac{(2\pi)^{d}(d-1)(d-2)(m^2+1)^{3}}{4g^2\Omega_d(m^2-1)}\right]^{-1}\!\!,
\end{equation}
which entails
\begin{equation}
g_*^2=\frac{(2\pi)^d (d-1)(d-2)(4-d)}{4\Omega_d(4d-7)}\,,
\end{equation}
providing  the non-Gaussian exponent
\begin{equation}
\omega = \epsilon-\frac{\epsilon^2}{3(3-\epsilon)}.
\end{equation}
Thus, in three dimensions $\omega=5/6$.

\section{The running Kähler potential}
\label{sec:Field-dependent Z section}

In a next step we incorporate the Kähler potential into our truncation:
\begin{align}
\begin{split}
\Gamma_k=&-\frac{Z_0}{4}\int\!\on{d}^d\!x\on{d}^2\!\theta\on{d}^2\!\bar{\theta}\;K(\Phi,\Phi^\dagger)\\
&-\left\{\frac{1}{2i}\int\!\on{d}^d\!x\on{d}^2\!\theta\; W(\Phi)+\on{h.\!c.}\right\}
\label{eq:truncKaehler}
\end{split}
\end{align}
The details on how to
extract the flow of the Kähler metric $\zeta(\Phi,\Phi^\dagger)=\partial_\Phi\partial_{\Phi^\dagger}K(\Phi,\Phi^\dagger)$
from the 
FRG equation~\eqref{eq:Wetterich}, as well as the most general result, are presented in App.~\ref{sec:appKaehler}.
Choosing $r_1=0$ we retain
\begin{align}
\Big(
\partial_t & 
-\Delta \left(\chi\partial_\chi
+\chi^\dagger\partial_{\chi^\dagger}\right)-\eta\Big)\zeta
=\int_q\!\frac{(\partial_t-q\partial_q-\eta)r_2}{v^3}
\nonumber\\
&\bigg\{ hu|w '''|^2-(u-2q^2h^2)\left(w ''^\dagger w '''\partial_{\chi^\dagger}\zeta+\operatorname{h.c.}\right)\nonumber\\
&\ \ -2q^2h(2|w ''|^2+u)|\partial_\chi\zeta|^2+uv\partial_\chi\partial_{\chi^\dagger}\zeta\bigg\}
\label{eq:zphiflow}
\end{align}
with $\chi$-dependent generalizations of the objects
in (\ref{eq:lpa_notations_uv}):
\begin{equation}
h=\zeta+r_2,\;\;u=|w ''|^2-q^2h^2,\;\;v=|w ''|^2+q^2h^2.
\end{equation}
Remarkably, for $w(\chi)$ being a monomial \Eqref{eq:zphiflow} admits the ansatz $\zeta(\rho=\chi\chi^\dagger)$. A cubic superpotential 
in $d=3$ yields
\begin{align}
\Big(\partial_t&-(1+\eta)\rho\partial_\rho-\eta\Big)\zeta=\int\limits_0^1\!\operatorname{d}\!q\;q^2\frac{(\partial_t-q\partial_q-\eta)r_2}{2\pi^2\, v^3}\nonumber\\
&\hphantom{=}\bigg\{
4hu\,g^2-8(u-2q^2h^2)g^2\rho\partial_\rho\zeta
\label{eq:zrhoflow}\\
&\hphantom{=}-2q^2h(8g^2\rho+u)\rho(\partial_\rho\zeta)^2+uv\left(\partial_\rho+\rho\partial_\rho^2\right)\zeta\bigg\}\nonumber
\end{align}
where, correspondingly:
\begin{equation}
u=4g^2\rho- q^2h^2\, ,\quad\quad v=4g^2\rho + q^2h^2\, .
\end{equation}
The occurrence of $\eta$ on the right-hand side of~\Eqref{eq:zrhoflow}
is a consequence of the RG-improvement of regulators in~\Eqref{eq:rescaling_regulators},
which is tantamount to requiring a deformation of the renormalized (instead
of bare) two-point-function. Though this accounts
for the resummations of perturbative contributions, it also
complicates considerably the structure of the flow equation.
Therefore, in the present study we confine ourselves to the approximation where such
contributions are neglected, thus effectively setting $\eta=0$ on 
the right-hand side of~\Eqref{eq:zrhoflow}.

Throughout this section we
adopt the Litim-type regulator I, augmented by a positive prefactor, 
\begin{equation}
r_2=a\left(\frac{1}{q}-1\right)\Theta(1-q^2),\;\;a\in\mathbb{R}^+.
\label{eq:litimrega}
\end{equation}
The parameter $a$ allows for a minimal sensitivity optimization: Results provided by truncated flow equations can be improved by minimizing their regulator-dependence~\cite{Liao:1999sh}. We implement this idea by targeting a stationary point of $\omega(a)$, since it is the less irrelevant critical
exponent that acquires a spurious regulator dependence within our truncations. 

\subsection{Fixed-point Kähler potential}
\label{sec:FixedPointPotential}

We consider the $\mathfrak{n}=3$ fixed point in three dimensions. 
Setting $\partial_t\zeta=0$ in  (\ref{eq:zrhoflow}) and $\eta_*=1/3$
on the left-hand side provides the fixed-point equation
\begin{align}
&\frac{1}{3}\zeta+\frac{4}{3}\rho\,\zeta'
+\frac{a}{2\pi^2}\int\limits_{0}^{1}\!\on{d}\!q\;\frac{q}{v^3}
\bigg\{ 4hug^2-8(u-2q^2h^2)g^2\rho\zeta'\nonumber\\
&\hphantom{0=}-2q^2h(8g^2\rho+u)\rho\,\zeta'^2+uv(\zeta'+\rho\zeta'')\bigg\}
=0\, .
\label{eq:zFlow2}
\end{align}
The integral can be solved analytically.

There are several methods at hand to analyze \eqref{eq:zFlow2}. We start with a polynomial truncation of the Kähler metric:
\begin{equation}
\zeta(\rho)=1+\sum\limits_{n=1}^{N}\zeta_{n}\rho^n\,
\label{eq:poltruncKaehler}
\end{equation}
The fixed-point equation \eqref{eq:zFlow2} is fulfilled to order $N$ if all its projections onto $\rho^n$, $n\leq N$ hold.
The resulting system of $N+1$ equations provides numerous roots. Yet, in all considered cases requiring the couplings to be real has left us with a unique solution for $(g_*^2,\zeta_{n*})$. 
In App.~\ref{sec:appKaehler} we provide the fixed point couplings obtained for $0\leq N\leq 5$ at several values of $a$.

\begin{figure}[!t] 
	\includegraphics[width=0.45\textwidth]{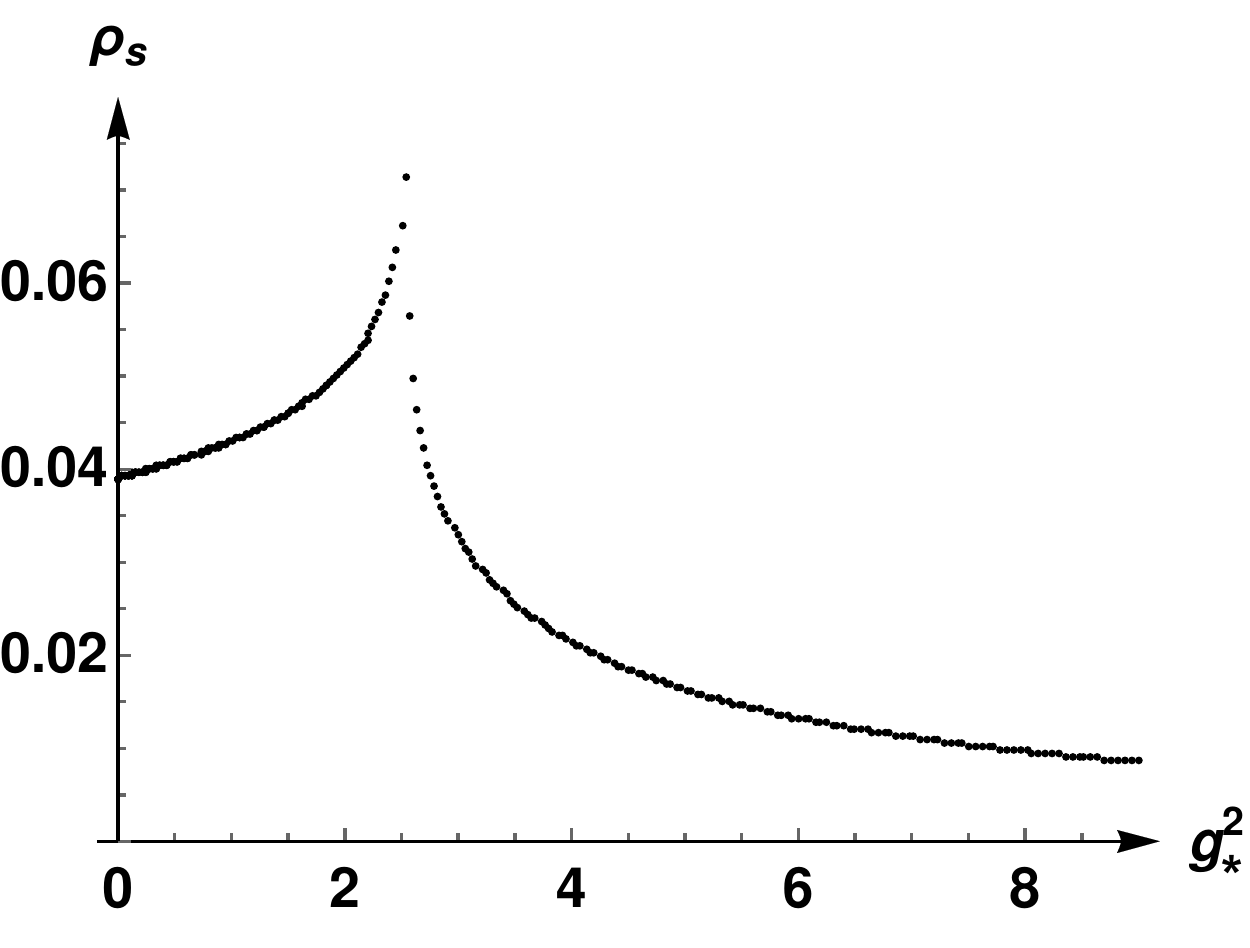}
	\caption{Shooting from the origin. Field coordinate $\rho_s$ of the singularity in $\zeta(\rho)$ closest to the origin as a function of $g^2$. Here $a = \num{1.7}$; the position of the spike provides $g_m^2(a=\num{1.7})=\num{2.53}$.}
	\label{fig:spikeplot}
\end{figure}

\begin{figure*}[t!] 
	\includegraphics[width=0.45\textwidth]{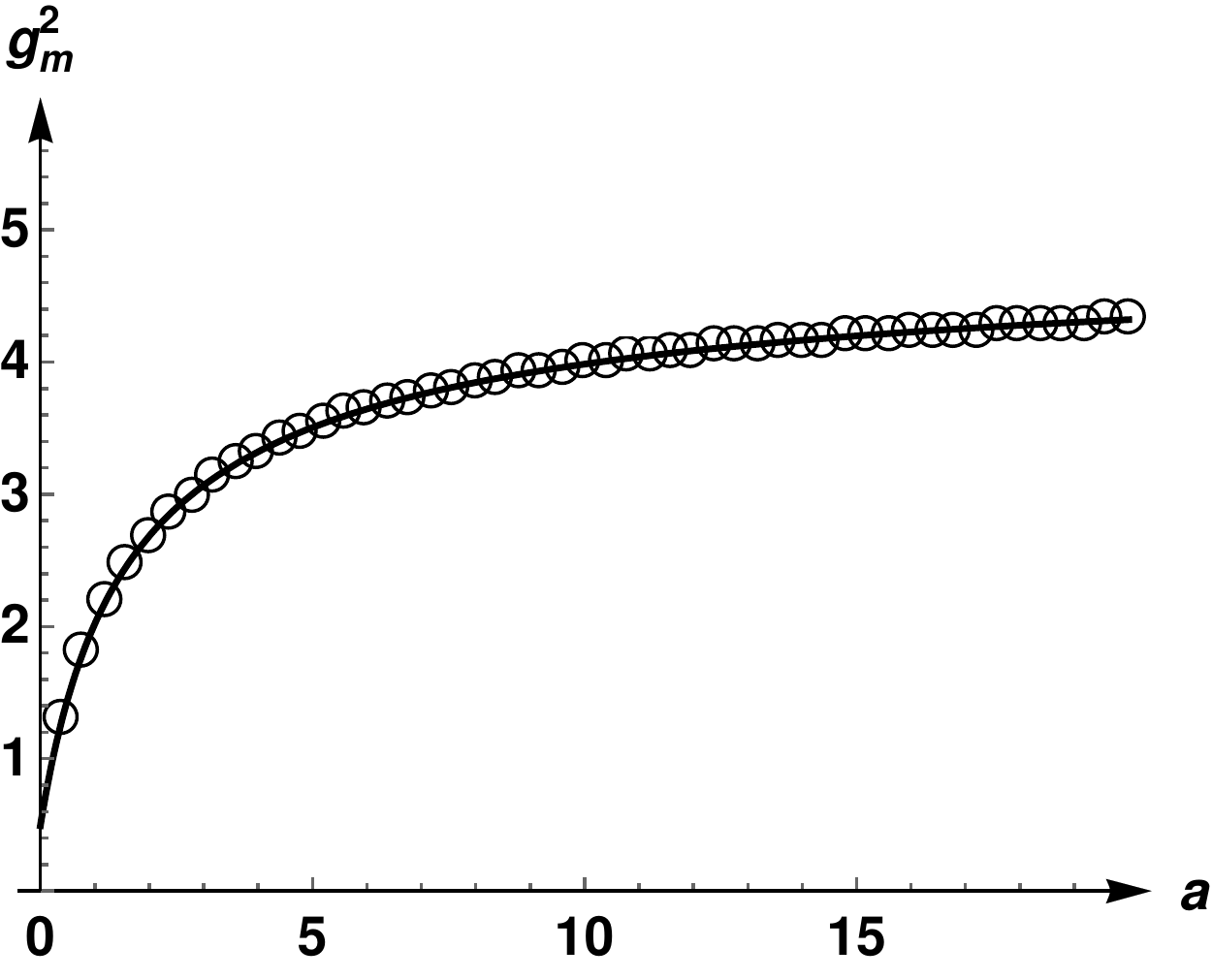}
	\hskip2mm
	\includegraphics[width=0.45\textwidth]{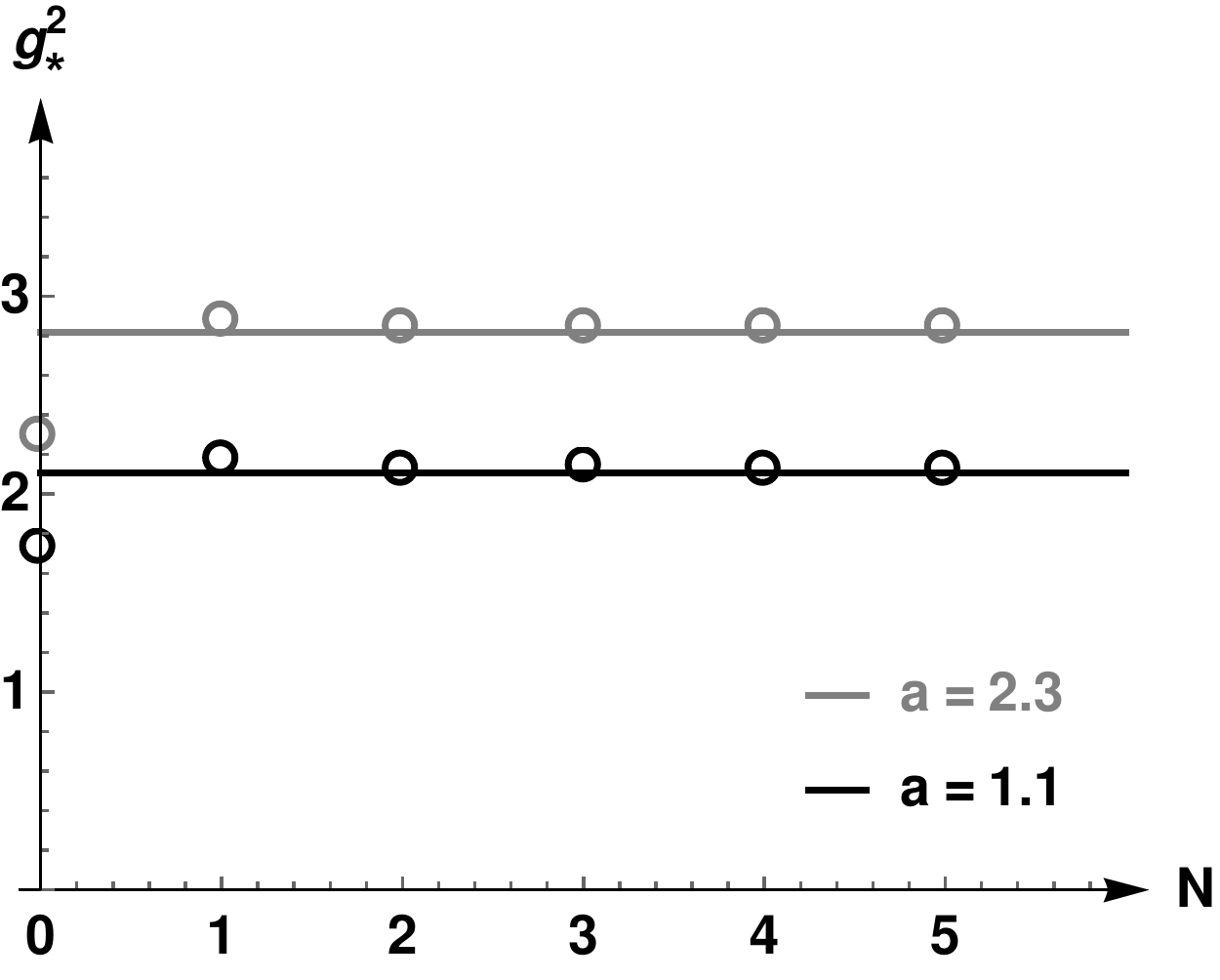}
	\caption{\emph{Left panel:} Fixed point values $g_m^2$ as provided by shooting from the origin for different $a$. The solid line is a fit; for the interpolating function see Eq.~\eqref{eq:fpshootingip}. \emph{Right panel:} Comparison of shooting from the origin and polynomial truncation. The solid line is located at $g_m^2(a)$ as obtained from the interpolating function. The $g_*^2(a,N)$ due to polynomial truncation of order $N$ converge to $g_m^2(a)$.}
	\label{fig:fpshooting}
\end{figure*}

To go beyond polynomial truncations we first employ 
numerical integration by shooting from the 
origin. For each $g^2$ and $a$, the fixed-point equation \eqref{eq:zFlow2} 
is a second-order nonlinear ordinary differential equation for 
$\zeta(\rho)$; hence we need to provide two initial conditions. 
Since we look for a solution that is smooth at the origin, 
the product $\rho\,\zeta''$ must vanish at $\rho=0$,
which gives a closed relation between $g$ and $\zeta'(0)$.
Therefore, while one condition is determined by $\zeta(0)=1$,
the second one, say $\zeta'(0)$, can be parametrized by $g^2$.
Yet, the normal form of \Eqref{eq:zFlow2} presents a
$1/\rho\,$-pole at the origin, which we avoid by 
imposing our regular initial conditions at
$\rho=\varepsilon\ll 1$.

Integrating \Eqref{eq:zFlow2} from
$\rho=\varepsilon$ outwards, we constantly hit a movable singularity. 
As illustrated by Fig.~\ref{fig:spikeplot}, the position $\rho_s(g^2)$ of this singularity exhibits a sharp maximum. Its location $g_m^2(a)$
is expected to correspond to the regular and polynomially bounded
solution of the truncated 
fixed-point equation \cite{Hasenfratz:1985dm,Morris:1994ki,Hellwig:2015woa}.
The left panel of Fig.~\ref{fig:fpshooting}
shows the dependence of $g_m^2$ on $a$. We interpolate it
using the fit
\begin{equation}
g_m^2(a)=\frac{\num{4.7951}a^2 + \num{31.796}a - \num{5.2531}}{a^2 + \num{9.0848}a - \num{10.624}}\, .
\label{eq:fpshootingip}
\end{equation}
The right panel of Fig.~\ref{fig:fpshooting} illustrates how the fixed-point values $g_*^2(a,N)$ obtained from the polynomial truncation~\eqref{eq:poltruncKaehler} converge to~$g_m^2(a)$.

Although shooting from the origin successfully predicts the unique 
critical $g_*^2$, it fails in producing a fixed-point K\"ahler metric
which is \emph{globally} defined in field space. The same applies to the polynomial
truncation of \Eqref{eq:poltruncKaehler},  since it likewise 
represents an expansion 
about the origin. To obtain the global critical
K\"ahler metric we employ pseudo-spectral methods,
 which are based on the expansion of $\zeta(\rho)$
 in a basis of Chebyshev polynomials
 (see Refs.~\cite{Borchardt:2015rxa,Borchardt:2016pif} for applications
 to FRG equations).
 Though this is again a polynomial expansion,
 we derive the system of corresponding fixed-point 
 equations not by a projection on the basis functions,
 but rather through a collocation method.
To this end, it is convenient to map the
$\rho$-domain into the compact interval $[0,1]$.
Using a Gauss-grid in this interval, it is then possible
	to adopt a numerical relaxation  method, such as for instance Newton-Raphson,
	to deduce the coefficients of $\zeta_*$ in the 
	Chebyshev basis.
Relaxation needs an initial seed, which is based on the information
obtained with the polynomial and shooting methods.
	
\begin{figure}[!t] 
	\includegraphics[width=0.45\textwidth]{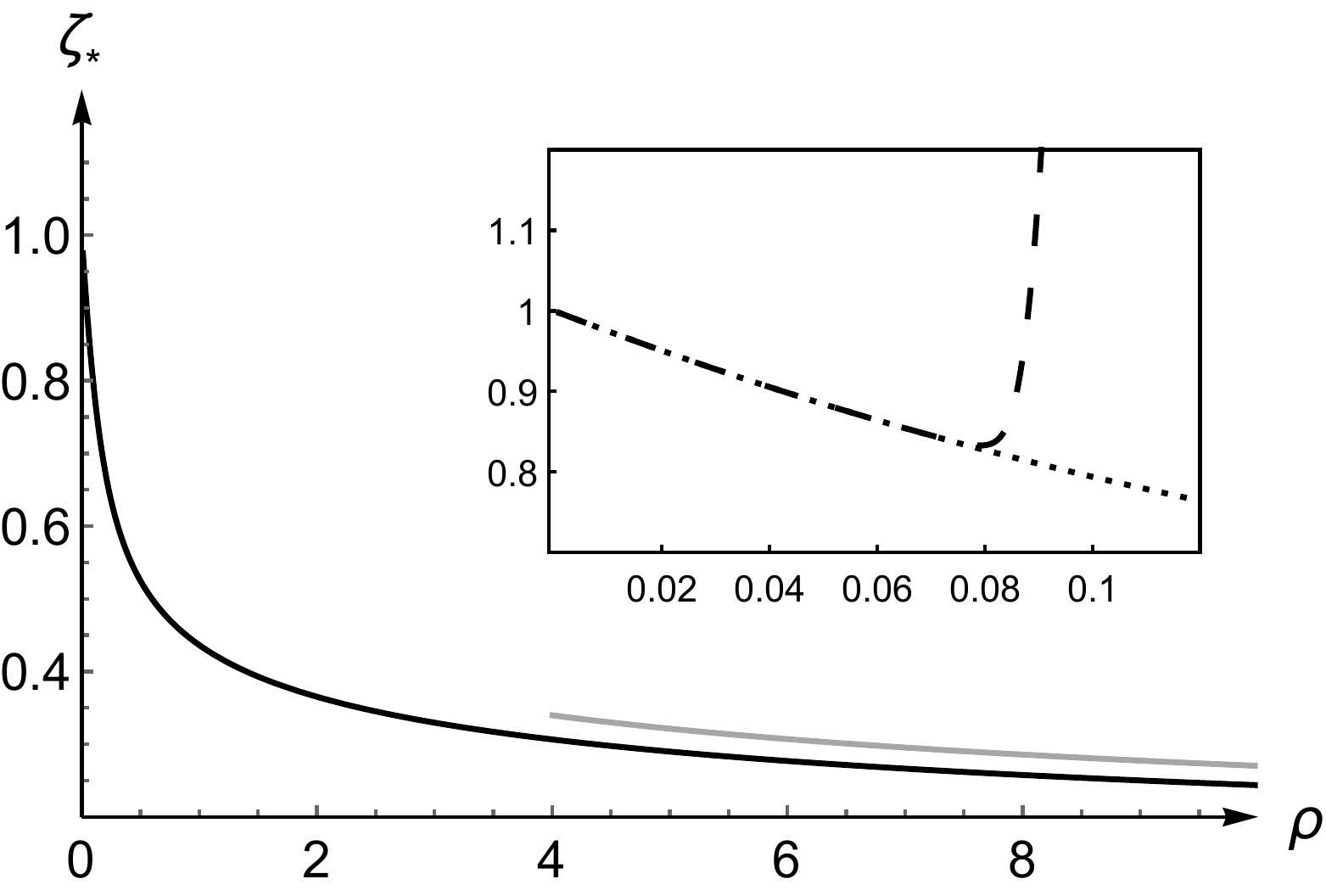}
	\caption{The critical K\"ahler metric
		for $a=\num{1.75}$. The solid black line is produced by pseudo-spectral methods,
		while the solid gray line shows a $\rho^{-1/4}$ decay.
		\emph{Inset:}
		 The pseudo-spectral solution (dotted line) is compared to the one obtained
		 by shooting from the origin (dashed line) with $g_*^2=\num{2.576}$
		 corresponding to a maximum of $\rho_s$ as in Fig.~\ref{fig:spikeplot}.}
	\label{fig:criticalzeta}
\end{figure}

The result of this analysis is presented in Fig.~\ref{fig:criticalzeta}.
It provides a smooth and featureless interpolation between the small-$\rho$ regime,
which is satisfactorily described by the polynomial truncations and the shooting
from the origin, and the large-$\rho$ region, where the K\"ahler metric is asymptotic to
$\rho^{-\eta_*/(2\Delta_*)}=\rho^{-1/4}$.
	
\subsection{Critical exponent {\boldmath$\omega$}}
\label{sec:critical_exponent_omega}

By tracking the RG flow of the K\"ahler metric,
	that is of infinitely many couplings which we
	expect to be irrelevant at the $\mathfrak{n}=3$
	fixed point, we can extract estimates of many more
	universal quantities, i.\,e. eigenvalues of the linearized
	RG equations, which are related to correction-to-scaling
	exponents. As a case-study, we focus on $\omega$ to test 
	the quality of different approximations.

We again start with a polynomial truncation.
For generic perturbations of the fixed-point superpotential,
which result in a $w_*+\delta w$ that is no longer a simple monomial,
the K\"ahler metric is no longer a function of the single invariant $\rho$.
Thus, in the polynomial truncation we refrain from combining $\chi$ and $\chi^\dagger$ into $\rho$:
\begin{align}
\label{eq:poltrunc2Kaehler}
\zeta(\chi,\chi^\dagger)=& \;1+\sum\limits_{n=1}^{N}\zeta_n(\chi\chi^\dagger)^n\\
&+\sum\limits_{n=0}^{N}\sum\limits_{m=1+n}^{2N-n}(\zeta_{nm}\chi^n\chi^{\dagger m}+\zeta_{mn}\chi^m\chi^{\dagger n})
\nonumber 
\end{align}
with $\zeta_n^\dagger=\zeta_n$ and $\zeta_{mn}^\dagger=\zeta_{nm}$. 
The corresponding stability matrix conveys the linearized flow equations of both the superpotential and the Kähler metric. 
Hence it accounts for the couplings $c_n$, appearing in the superpotential, as well as $\zeta_n$ and $\zeta_{nm}$. 
For $r_1=0$, the stability matrix becomes block diagonal such that $g$ is coupled solely to $\{\zeta_n\}$. 
Restricting ourselves to this submatrix we can resort to the simplified flow equation \eqref{eq:zrhoflow} for $\zeta(\rho)$, 
still for $\eta=0$ on its right-hand side and the Litim-type regulator I from \Eqref{eq:litimrega}.

\begin{table}[t!]
	\begin{tabular}{l|rrrrrrr}
		$N$ &\multicolumn{1}{c}{$a=\num{1.1}$} &\multicolumn{1}{c}{$a=\num{1.3}$} &\multicolumn{1}{c}{$a=\num{1.5}$} &\multicolumn{1}{c}{$a=\num{1.7}$} &\multicolumn{1}{c}{$a=\num{1.9}$} &\multicolumn{1}{c}{$a=\num{2.1}$} &\multicolumn{1}{c}{$a=\num{2.3}$}    \\\hline
		0 & \num{1.0000} &  \num{1.0000} &  \num{1.0000}  & \num{1.0000} & \num{1.0000} & \num{1.0000} & \num{1.0000} \\
		1 & \num{0.8270} &  \num{0.8370} &  \num{0.8422}  & \num{0.8446} & \num{0.8454} & \num{0.8452} & \num{0.8444} \\
		2 & \num{0.8292} &  \num{0.8310}& \num{0.8317} & \num{0.8317} & \num{0.8309} & \num{0.8298} & \num{0.8284}  \\
		3 & \num{0.8258} &  \num{0.8307} & \num{0.8330} & \num{0.8338} & \num{0.8338} & \num{0.8331} & \num{0.8321} \\
		4 & \num{0.8279} &  \num{0.8319} & \num{0.8338} & \num{0.8345} & \num{0.8343} & \num{0.8336} & \num{0.8326}
	\end{tabular}
\caption{Critical exponent $\omega$ due to polynomial truncation of $\zeta(\rho)$ to order $N$. \label{tab:omegas1}}
\end{table}

Identifying the smallest positive eigenvalue of the stability submatrix with $\omega$ we obtain the results presented in Tab.~\ref{tab:omegas1}.
We have verified that all corresponding eigendirections indeed provide a non-vanishing $\delta g$.
The computation of $\omega$ has turned out to be very memory-consuming. This has limited us in both the achievable truncation order $N$ and the step size of $a$. 
Yet we observe that, for all $N$,  
$\omega(a)$ exhibits a stationary point, whose position $a_s$ 
depends on $N$. At forth order the best approximation we have obtained is 
$\omega=\num{0.8345}$ at $a_s=\num{1.7}$. Convergence is achieved only for the first two significant figures.

Since polynomial truncations are computationally demanding, 
also for the determination of $\omega$ we again make use of shooting from the origin.
Just as we linearized the flow equation of the superpotential in
the fluctuations $\delta w=w-w_*$ and
$\delta \eta=\eta-\eta_*$, obtaining \Eqref{eq:linwflow},
we can also linearize the flow of the K\"ahler metric
with respect to $\delta w$, $\delta \eta$ and $\delta \zeta(\rho)=\zeta(\rho)-\zeta_*(\rho)$ to obtain a second-order linear partial differential equation.
The eigendirections with $\delta w_\lambda\neq 0$, $\delta\zeta_{\lambda}=0$
are the ones already described in \Eqref{eq:linsoldw},
for which no knowledge of $\delta\eta$ is required. 
The remaining perturbations with $\delta\zeta_{\lambda}\neq0$ instead depend on 
$\delta\eta$, and according to our definitions must have  $\delta\zeta(0)=0$.
%
%
To extract $\omega$ we look for eigensolutions 
with $\lambda=\omega$ and
\begin{equation}
\delta w_\omega(\chi)=\frac{\delta g}{3}\, \chi^3.
\label{eq:deltaw_omega}
\end{equation}
Thus \Eqref{eq:linwflow} and \Eqref{eq:eigenvalueproblem} can be analytically solved
for $\delta g$ as a function of two unknowns: $\omega$ and $\delta \eta$.
Then, the eigenvalue problem reduces to the solution of the single 
second-order ordinary differential equation 
for $\delta\zeta_\omega$,
involving the two free parameters $\omega$ and $\delta \eta$.
In addition, two initial
conditions have to be supplied to specify a unique solution. 
We choose to provide them at the origin.
As in the fixed point case, requiring the solution 
to be smooth at $\rho=0$ together with consistency
between the initial conditions and the differential equation
itself determines one of these conditions,
say $\delta \zeta_\omega'(0)$.
Since the other one is provided by $\delta\zeta_\omega(0)=0$, 
the space of eigenfunctions is completely spanned by
$\omega$ and $\delta \eta$.

While the former becomes quantized, by a mechanism that will be
	explained in the following, the latter remains a free parameter, 
	as in the first order of the $\epsilon$ expansion, see Sec.~\ref{sec:LPA$'$ section}.
 Indeed, the considered differential equation is
linear, such that the overall normalization of $\delta\zeta_\omega$ 
is arbitrary, and does not play a role in the determination
of $\omega$. 
So, if the choice of $\delta\eta$ affects only the
overall normalization of $\delta\zeta_\omega$,
different values of $\omega$ correspond to
genuinely different solutions. 
In particular, by inspecting the behavior of such
solutions, one finds that they rapidly grow,
with a rate that appears to be exponential.  
Indeed, these exponential parts are almost invariably present
in solutions of linearized FRG equations. It is the
additional requirement that the eigenfunctions $\delta\zeta_\lambda$
have to follow a power law for large $\rho$
that quantizes the set of possible eigenvalues.
This requirement in turn is related to 
self-similarity~\cite{Morris:1998da} and to the existence of a
well-defined norm in theory space~\cite{Bridle:2016nsu,ODwyer:2007brp}.

Thus, we expect a unique value of $\omega$ corresponding
to a $\delta\zeta_\omega$ which is asymptotic to some power of $\rho$
for large $\rho$. 
In practice, a simple way to determine such a value consists
in plotting $\delta\zeta_\omega(\rho_L)$ as a function of
$\omega$ for large enough $\rho_L$.
Then, the solution with power-law asymptotic behavior
should correspond to a special $\omega$ such that
$\delta\zeta_\omega(\rho_L)$
is exponentially smaller than for all other values outside
a small neighborhood of it. 
We apply this criterion to the solutions constructed
by shooting from the origin. 
In this case, each fixed-point solution extends over a 
finite range $[0,\rho_s]$. Therefore we
parametrize $\rho_L=(1-\delta)\rho_s$,
and scan over $\delta\ll 1$. 
We observe that $\delta\zeta_\omega(\rho_L)$
as a function of $\omega$
shows only one zero, 
where it changes sign. 
This change of sign can be made arbitrarily quick 
by choosing smaller and smaller values of $\delta$. 
Its location
converges to an unambiguous value in the limit
$\delta\to0$. 
This zero can be identified with the physical
value of $\omega$.
In the interval $a\in[\num{1.6},\num{1.9}]$  we find $\omega=\num{0.834}$,
with variations only in the fourth decimal place, showing a maximum at approximately $a=\num{1.8}$, where $\omega=\num{0.8344}$.

\section{Momentum-dependent K\"ahler potential}
\label{sec:momentumdependent Z section}

Another branch of possible truncations is offered by the generalized Kähler potential with minimal field content:
\begin{align}
\begin{split}
\Gamma_k=&-\frac{Z_0}{4}\int\!\on{d}^d\!x\on{d}^2\!\theta\on{d}^2\!\bar{\theta}\;\Phi^\dagger z(D,\bar{D})\Phi\\&-\left\{\frac{1}{2i}\int\!\on{d}^d\!x\on{d}^2\!\theta\; W(\Phi)+\on{h.\!c.}\right\}
\label{eq:ansatzMomentum}
\end{split}
\end{align}
with analytic and Hermitian $z(D,\bar{D})$ fulfilling $z(0,0)=1$. 
Just as for the regulator functions $\rho_2$ and $r_2$ in (\ref{eq:DeltaS1}) and (\ref{eq:DeltaS2}), $z$ can be replaced by an analytic Hermitian generalized Kähler metric $\zeta(-\partial_x^2)$ with $\zeta(0)=1$.

The flow of $\zeta$ can be obtained from the functional 
$\phi\,$- and $\phi^\dagger$-\,derivative of the FRG equation~\eqref{eq:Wetterich} at vanishing fields. $\Gamma_k^{(2)}(p,q)$ at constant $f$ and zero $\psi$, $\bar{\psi}$ is provided in App.~\ref{sec:appMomentum}. Since it is no longer proportional to $\delta(p-q)$, $(\Gamma_k^{(2)}+R_k)^{-1}$ cannot be computed just by matrix inversion. Hence, to evaluate the projection we proceed as described for instance in Ref.~\cite{gieswetterich}: We rewrite the flow equation \eqref{eq:Wetterich} as
\begin{equation}
\partial_t \Gamma_k=\frac{1}{2}\on{STr}\left(\tilde{\partial}_t \ln(\Gamma_k^{(2)}+R_k)\right),
\end{equation}
where $\tilde{\partial}_t$ is assumed to act on $R_k$ only, and expand the logarithm about the field independent part $\Gamma_0\propto \delta(p-q)$ of $\Gamma_k^{(2)}+R_k=:\Gamma_0+\Delta \Gamma$,
\begin{align}
\begin{split}
& \ln(\Gamma_k^{(2)}+R_k)=\ln(\Gamma_0)+\Gamma_0^{-1}\Delta\Gamma-\frac{1}{2}(\Gamma_0^{-1}\Delta\Gamma)^2+\ldots\, .
\end{split}
\end{align}
Because every non-vanishing entry in $\Delta\Gamma$ is at least linear in $\phi$ or $\phi^\dagger$, we have to consider only addends containing $\Delta\Gamma$ once or twice.

The dimensionless, renormalized flow equation for $\zeta$ amounts to
\begin{align}
&(\partial_t -p\partial_p-\eta)\zeta(p^2)=-\int\!\!\frac{\on{d}^d\!q}{(2\pi)^d}\;\frac{4g^2h(q-p)}{v^2(q)v(q-p)}\times  
\label{eq:dlessflowBLPA1}\\
&\times \left[2hM(\partial_t-q\partial_q-\eta+1)r_1
-u(\partial_t-q\partial_q-\eta)r_2\right](q)\, ,\nonumber
\end{align}
where all functions in the second line are evaluated at~$q$.
$M$, $u$ and $v$ are defined in~\Eqref{eq:lpa_notations_uv}
and $h=\zeta+r_2$.
Discarding the momentum dependence of $\zeta$ 
in \Eqref{eq:dlessflowBLPA1} 
restores the LPA$'$ result of \Eqref{eq:flowlpa}. At $d=2$ we recover the flow equation for the two-dimensional $\mathcal{N}=(2,2)$ WZ model  derived in Ref.~\cite{Synatschke:2010jn}. The apparent
discrepancy by a factor of two originates from a difference in the definitions of coupling constants and regulator functions. The flow equation \eqref{eq:dlessflowBLPA1} implies $\eta\propto g^2$, as becomes obvious when setting $p=0$. Hence, just as the LPA$'$ truncation, it accounts only for the 
Gaussian and $\mathfrak{n}=3$ fixed points.

In the present paper we confine ourselves to the first-order polynomial truncation of the Kähler metric,
\begin{equation}
	\zeta(p^2)= 1 + p^2\zeta_1.
\end{equation}
Thus, we have to consider the projections of Eq.~\eqref{eq:dlessflowBLPA1} onto its zeroth and first orders in $p^2$. We report on the results obtained 
for the three-dimensional $\mathfrak{n}=3$ fixed point
by adopting three different regulators.
All arising integrals have been solved analytically.
A further exploration of the ansatz in \Eqref{eq:ansatzMomentum} is yet to be addressed.


We start with the Callan-Symanzik regulator
	defined in~\Eqref{eq:CallanSymanzik}.
Examining the vicinity of $\zeta_1=0$ at $\eta_*=1/3$ we numerically find the fixed point solution
\begin{equation}
	(\zeta_{1*},\,g_*^2)=(\num{-0.0816},\,\num{2.9425}).
\end{equation}
The corresponding stability matrix couples $g$ to $c_2$ and $\zeta_1$. Its spectrum consists of \Eqref{eq:linsoldw} supplemented by the two eigenvalues
\begin{equation}
	\omega = \num{0.6687},\;\; \lambda_\zeta=\num{-75.75}.
\end{equation}
We expect the additional relevant exponent $\lambda_\zeta$
 to be an error induced by the 
combined effect of truncation and regularization scheme.
This is supported by the results obtained with the other two regulators.


Despite its discontinuity in momentum space, the Litim-type regulator I, see \Eqref{eq:litimreg}, provides a finite flow of $\zeta_1$.
The arising integrals converge for $\zeta_1>-1$. Note, however, that applying step-like regulators to higher orders of a $p^2$ expansion is problematic~\cite{Bonanno:1999ik,Canet:2003qd}.
Setting $\eta_*=1/3$ we numerically obtain the fixed point values
\begin{equation}
	(\zeta_{1*}, \,g_*^2)=(\num{-0.0138},\,\num{1.9509}).
\end{equation}
The stability matrix couples $g$ to $\zeta_1$. The corresponding eigenvalues evaluate to
\begin{equation}
\omega=\num{0.8317},\;\;\lambda_\zeta=\num{2.530}.
\label{eq:lambdazeta1}
\end{equation}


To further simplify the flow equation 
we turn to another step-wise regulator,
which we call the Litim-type regulator II.
We set
\begin{equation}
r_1=0,\;\;r_2=\zeta(q^2)\left(\frac{1}{q}\frac{\zeta(1)}{\zeta(q^2)}-1\right)\Theta(1-q^2).
\label{eq:litimreg2}
\end{equation}
For $\zeta=1$ this coincides with our definition of Litim-type regulator I. 
With \Eqref{eq:litimreg2} again only low-energy modes, $q<1$, contribute to the
flow, going along with $h(q)=\zeta(1)/q$ such that $u$ and $v$ become momentum-independent. Let us stress that, contrary to the other regulators adopted in this
work, for the present choice the $\partial_t$-derivative on the right-hand side of
\Eqref{eq:dlessflowBLPA1} gives a non-vanishing contribution. The fixed point equation becomes
\begin{align}
\begin{split}	
&35\zeta_{1*}(2+3\zeta_{1*})(1+\zeta_{1*})+(1+3\zeta_{1*})^2=0\, ,\\
&g_*^2=\frac{2\pi^2}{5}\frac{(1+\zeta_{1*})^3}{2+3\zeta_{1*}}\,
\end{split}
\end{align}
yielding three solutions.
It is a common feature of polynomial truncations to suggest spurious fixed points. A comparison with our previous findings allows to identify the physical result as
\begin{align}
\begin{split}
g_*^2&=\num{1.9339}\, ,\quad\quad
\zeta_{1*}=\num{-0.0136}\, ,\\
\omega&=\num{0.8443}\, ,\quad\quad\,\,
\lambda_\zeta=\num{2.411}\, .
\label{eq:lambdazeta2}
\end{split}
\end{align}

The good agreement between \Eqref{eq:lambdazeta1} and \Eqref{eq:lambdazeta2}
suggests that we could qualitatively trust also the estimate of $\lambda_\zeta$.
Finally, let us remark that the negative sign of $\zeta_{1*}$ does not 
necessarily signal the presence of negative norm states.
In fact, within a polynomial truncation of $\zeta(p^2)$
couplings with alternating signs are allowed and might 
be needed for convergence of the power series.

\section{Multicritical models}
\label{sec:Multicrit}

The critical models with $\mathfrak{n}>2$
can be constructed by using perturbation theory
in vicinity of the upper critical dimensions $d_\mathfrak{n}$ defined in~\Eqref{eq:upper_critical_d},
where they are weakly coupled.
In fact, if $d=d_\mathfrak{n}-\epsilon$
the constraint from~\Eqref{eq:etas}
entails
\begin{equation}
\eta_*=\epsilon\frac{\mathfrak{n}-2}{\mathfrak{n}}\, .
\label{eq:eta_epsilon}
\end{equation}
This allows for a standard $\epsilon$ expansion in the spirit
of Ref.~\cite{Wilson:1973jj}.
This approach has already been applied to non-supersymmetric multicritical models
in fractional dimensions~\cite{Itzykson:1989sx}.
Some of these studies have been performed directly
in a FRG setup~\cite{Nicoll:1974zz,ODwyer:2007brp,Codello:2017hhh}.

For infinitesimal $\epsilon$ the deviations of the fixed-point
values $w_*$ and $\zeta_*$ from the Gaussian ones $w_*=0$ and $\zeta_*=1$,
as well as those of the eigenperturbations $\delta w$ and $\delta \zeta$
from the corresponding Gaussian eigenperturbations, behave as
positive powers of $\epsilon$. 
Since the term $\eta\zeta$ on the left-hand side 
of~\Eqref{eq:zphiflow} is of order $\epsilon$,
we assume that the same holds for the right-hand side
of this equation, i.\,e. we do not consider possible solutions with
$(\zeta-1)$ of order $\epsilon^P$ with $P<1$.
Instead, we assume that the power-counting in $\epsilon$ which we discussed for
the $\mathfrak{n}=3$ model in Sec.~\ref{sec:LPA$'$ section}
applies to any value of $\mathfrak{n}$.

In this case, the lowest order in~\Eqref{eq:zphiflow} is of order $\epsilon$. This requires that $|c_{\mathfrak{n}*}|^2\propto\epsilon$
and, for $\mathfrak{n}>3$, also $\zeta'\propto\epsilon$.
It follows that at leading order the flow equation~\eqref{eq:zphiflow}
 becomes
\begin{equation}
\Big(\partial_t 
-\Delta_\mathfrak{n} \left(\chi\partial_\chi
+\chi^\dagger\partial_{\chi^\dagger}\right)
+l_{0\mathfrak{n}}\partial_\chi\partial_{\chi^\dagger}\Big)\zeta
=\eta-l_{1\mathfrak{n}} |w'''|^2\,,
\label{eq:leading_epsilon_flow}
\end{equation}
where $\Delta_\mathfrak{n}=(d_\mathfrak{n}-2)/2=1/(\mathfrak{n}-2)$ 
and we have defined the positive integrals
\begin{align}
\begin{split}
l_{0d}=\int_q\!\frac{(\partial_t-q\partial_q)r_2}{(1+r_2)^2 q^2}\, ,\\
l_{1d}=\int_q\!\frac{(\partial_t-q\partial_q)r_2}{(1+r_2)^3 q^4}\, ,
\end{split}\label{eq:l_integrals}
\end{align}
which are denoted as 
$l_{0\mathfrak{n}}$ and $l_{1\mathfrak{n}}$
when evaluated at $d=d_\mathfrak{n}$.
Let us stress that these numbers are, in general, regulator dependent.
Only $l_{1\mathfrak{n}=3}$ is universal, since it corresponds
to the one-loop anomalous dimension of the $\mathfrak{n}=3$ model.
For $\mathfrak{n}>3$  the leading order of the $\epsilon$ expansion
accounts for multi-loop diagrams.
The ansatz of~\Eqref{eq:truncKaehler} is one-loop exact
but it fails in reproducing all perturbative contributions
beyond one loop. Thus, the $\epsilon$
expansion of~\Eqref{eq:zphiflow} does not include all the contributions
to leading order in $\epsilon$, which explains the appearance of
nonuniversal coefficents.
One can nevertheless extract 
approximate results from truncated and perturbatively expanded
FRG equations, as is shown in Refs.~\cite{Nicoll:1974zz,ODwyer:2007brp}.

At the fixed point $|w_*'''|^2$ on the right-hand side of~\Eqref{eq:leading_epsilon_flow}
depends on $\rho$ only, such that 
~\Eqref{eq:leading_epsilon_flow}
allows for radial solutions $\zeta_*(\rho)$.
These are defined by a linear first-order 
ordinary differential equation for 
$\zeta_*'(\rho)$.
The physical solutions read
\begin{equation}
\zeta_*'(\rho)=\epsilon\frac{\mathfrak{n}-2}{\mathfrak{n}\, l_{0\mathfrak{n}}}
\sum_{i=0}^{\mathfrak{n}-4}\frac{1}{(i+1)!}\left(\frac{2\rho}{(\mathfrak{n}-2)l_{0\mathfrak{n}}}\right)^i\, .
\label{eq:multicriticalFP}
\end{equation}
The condition that $\zeta_*'(0)$ be finite 
requires the cancellation of $1/\rho$ poles
and fixes
the coupling $c_\mathfrak{n}$ to the value
\begin{equation}
c_{\mathfrak{n}*}^2=
\frac{\epsilon \ \left(l_{0\mathfrak{n}}/2\right)^{3-\mathfrak{n}}}{\mathfrak{n} (\mathfrak{n}-1)^2(\mathfrak{n}-2)^{\mathfrak{n}-2} (\mathfrak{n}-3)! \ l_{1\mathfrak{n}}}\, .
\label{eq:FPc_epsilon}
\end{equation}
This is universal only for $\mathfrak{n}=3$.
Since the radial fixed-point equation is a first-order ordinary differential equation,
its space of solutions is parametrized by one integration constant, which we did not discuss so far. Indeed one could add to 
\Eqref{eq:multicriticalFP} a term of the form 
\begin{equation}
 \frac{K}{\rho}e^{\frac{2\rho}{(\mathfrak{n}-2)l_{0\mathfrak{n}}}}\, .
 \label{eq:term_notto_add}
\end{equation}
The constant $K$ has to be set to zero, to ensure that  the space of
perturbations of the fixed point possesses a well-defined norm,
a countable basis  and a discrete spectrum~\cite{ODwyer:2007brp,Bridle:2016nsu}.

Once the particular solution in \Eqref{eq:multicriticalFP} is known,
it is possible to construct the general fixed-point solution through addition of the
solutions of the homogeneous part of \Eqref{eq:leading_epsilon_flow}.
The latter can be constructed by factoring the radial and the angular dependence,
as will be detailed for the linear eigenperturbations in the following.
The angular solutions are simple periodic functions 
$e^{i\mathfrak{m}\vartheta}$ labeled by the integer $\mathfrak{m}$.
For any non-vanishing $\mathfrak{m}$, the radial component of the 
homogeneous solutions
contains either a singularity at the origin or an exponentially growing part.
We discard such solutions and set $\mathfrak{m}=0$. 

Since at the upper critical dimension the fixed points are Gaussian,
the eigenvalue problem for the linearized flow in the $\epsilon$
expansion can be interpreted as a perturbation of the Gaussian case.
Therefore we first address the latter.

\subsection{Linearized flow at the Gaussian fixed point}
\label{sec:linearized_Gaussian}

For the free theory, with $w_*=0$, $\eta_*=0$, and $\zeta_*=1$,
the linearized flows of $\delta w$ and $\delta \zeta$ are decoupled,
since \Eqref{eq:linwflow} becomes independent of $\delta \eta$,
while for the K\"ahler metric one finds
\begin{equation}
\partial_t\delta\zeta=\delta\eta+
\left( \frac{d-2}{2}\left(\chi\partial_\chi
+\chi^\dagger\partial_{\chi^\dagger}\right)-l_{0d}\partial_\chi\partial_{\chi^\dagger}
\right)\delta\zeta\, .
\label{eq:zgausslinflow}
\end{equation}
Thus, there are separate families of eigendirections.
Members of the first family have a perturbed superpotential only:
\begin{align}
\begin{split}
\delta w_n&=(\delta c_n \chi^n)/n\,,\\
\lambda_n&=1-d+ n\, \frac{d-2}{2}\, ,
\label{eq:gausslinsoldw}
\end{split}
\end{align}
and $\delta \zeta_n=\delta \eta_n=0$.
Members of the second family have only a perturbed K\"ahler metric,
which is  conveniently expressed in spherical
coordinates
\begin{equation}
\Call{r}=\sqrt{\frac{\rho}{l_0}}\,,\quad
\vartheta=\arctan\left(i\,\frac{\chi^\dagger-\chi}{\chi^\dagger+\chi}\right)\, . 
\label{eq:radius_angle}
\end{equation}
The eigenvalue problem in these coordinates reads
\begin{equation}
\left[\lambda
-\frac{d-2}{2}\Call{r}\, \partial_{\!\call{r}}
+\frac{1}{4}\left(\partial_{\!\call{r}}^2+
\frac{1}{\!\Call{r}}\partial_{\!\call{r}}\right)
+\frac{1}{4\Call{r}^{\,2}}\partial_\vartheta^2
\right]\!\delta\zeta
=\delta\eta\, .
\label{eq:sphericalgausslinflow}
\end{equation}
Separable solutions can be obtained from the ansatz
\begin{equation}
\delta\zeta = \frac{\delta\eta}{\lambda}+e^{i m\vartheta}\,Q(\kern -0.9pt \Call{r}\,)
\end{equation}
giving rise to the following radial eigenvalue equation:
\begin{equation}
\left[\lambda-\frac{m^2}{4\Call{r}^{\,2}}
-\frac{d-2}{2}\Call{r}\, \partial_{\!\call{r}}
+\frac{1}{4}\left(\partial_{\!\call{r}}^2+
\frac{1}{\!\Call{r}}\partial_{\!\call{r}}\right)
\right]\!Q
=0\, 
\label{eq:radgausslinflow}
\end{equation}
Again, we constrain the space of solutions by prohibiting
singularities
at the origin and exponential growth for large radii.
This eliminates half of the solutions and quantizes $\lambda$
to the following discrete spectrum:
\begin{align}
\begin{split}
\lambda_{km}&=(d-2)\left(k+\frac{|m|}{2}\right)\, ,\quad 
k\in \mathbb{N}_0\,,\;\;
m\in \mathbb{Z}\, \text{ even}\\
Q_{km}&= b_{km} \binom{k+|m|}{k}^{-1}\!\! \left((d-2)\Call{r}^{\,2}\right)^{\frac{|m|}{2}}\mathrm{L}_k^{|m|}\!\left((d-2)\Call{r}^{\,2}\right),
\end{split}
\label{eq:gausslinsoldzeta}
\end{align}
where $\mathrm{L}$ denotes the generalized Laguerre polynomials and $b_{km}$ is an arbitrary normalization factor.
The condition $\delta\zeta(0)=0$ determines $\delta\eta$:
\begin{equation}
\delta\zeta(0)=\frac{\delta\eta}{\lambda_{km}}+b_{km}\delta_{m0}=0,
\label{eq:constraint_on_b}
\end{equation}
such that the perturbations with $m=0$ can have a non-vanishing $\delta\eta$,
which then scales with $b_{km}$.
For all other eigendirections with $m\neq 0$ we have $\delta\eta=0$.

For special values of $d$, which are precisely of the form of
$d_\mathfrak{n}$ in \Eqref{eq:upper_critical_d}, provided
\begin{equation}
n=\mathfrak{n}+2k+|m|\, ,
\end{equation}
the two distinct subspaces of eigensolutions contain degenerate solutions going along with the same eigenvalue.

\subsection{Critical exponent {\boldmath $\omega$} for general {\boldmath $\mathfrak{n}$}}
\label{sec:omega_multicritical}

Let us now turn to the problem of determining the critical exponents of the multicritical models away from their upper critical dimensions.
As the analysis in Sec.~\ref{sec:constraints} shows, the nonrenormalization of the superpotential
imposes quantization rules for the critical $\eta$
 and for the part of the spectrum described by \Eqref{eq:linsoldw}. 
These can be straightforwardly rewritten using $d=d_\mathfrak{n}-\epsilon$
and provide eigenvalues that are linear in $\epsilon$, since
$\Delta_*=\Delta_\mathfrak{n}-\epsilon/\mathfrak{n}$.
In particular, they support the expectation that the number of physically relevant directions
at the $\mathfrak{n}$-th fixed point be equal to $\mathfrak{n}-2$,
though they	 do not describe the classically marginal case $n=\mathfrak{n}$.
The latter has been already observed to become irrelevant for $\mathfrak{n}=3$ in the past sections.
We now adopt the $\epsilon$ expansion to address this computation for generic $\mathfrak{n}$.

Integration of \Eqref{eq:linwflow} for generic $\mathfrak{n}$ leads
to
\begin{equation}
\delta w\!\left(\chi\right)=\frac{c_{\mathfrak{n}*}\delta\eta}{2\lambda}
\chi^\mathfrak{n}
+\delta c_\lambda \chi^{\mathfrak{n}+\frac{\mathfrak{n}\lambda}{d-1}}\, .
\label{eq:deltaw_ngeneric}
\end{equation}
We focus on perturbations with $\delta c_\lambda=0$, which correspond to $\lambda=\omega$. 
As for the case $\mathfrak{n}=3$ discussed in Sec.~\ref{sec:LPA$'$ section},
to determine $\lambda$ additional knowledge from the running of $\zeta$ is needed.
Before moving to the latter, let us stress that 
\Eqref{eq:linsoldw} and \Eqref{eq:deltaw_ngeneric} can also be obtained
by the $\epsilon$ expansion of the eigenvalue problem with the ansatz
\begin{equation}
\begin{split}
\lambda&=
\epsilon\,\lambda^{(1)}\, ,\\
\delta\eta&=\sqrt{\epsilon}\, \delta\eta^{(1)}\, ,\\
\delta w&= \delta w^{(0)}+\epsilon\,\delta w^{(1)}\, .
\end{split}
\label{eq:epsilon_exp_perturbation_part1}
\end{equation}
At zeroth order in $\epsilon$ the Gaussian
solution goes along with the eigenvalue $\lambda^{(0)}=0$,
such that \Eqref{eq:deltaw_ngeneric} relates the zeroth-order 
superpotential to
the first-order eigenvalue by
\begin{equation}
\frac{\delta w_\mathfrak{n}^{(0)}}{\chi^\mathfrak{n}}=\frac{\delta c_\mathfrak{n}^{(0)}}{\mathfrak{n}}
=\frac{c_{\mathfrak{n}*}\delta\eta}{2\lambda}\, .
\label{eq:deltac_deltaeta}
\end{equation}
where the right-hand side has to be expanded at lowest order in $\epsilon$.

Let us then turn to the perturbation of
the K\"ahler metric.
We complement \Eqref{eq:epsilon_exp_perturbation_part1} with
\begin{equation}
\delta \zeta= 
\sqrt{\epsilon}\,\delta \zeta^{(1)}\, .
\label{eq:epsilon_exp_perturbation_part2}
\end{equation}
Thus the leading non-trivial contribution in
the $\epsilon$ expansion of the eigenvalue equation for $\delta\zeta$
is of order $\sqrt{\epsilon}$ and reads
\begin{align}
\begin{split}
\delta\eta \,+&
\left( \Delta_\mathfrak{n}\left(\chi\partial_\chi
+\chi^\dagger\partial_{\chi^\dagger}\right)-l_{0\mathfrak{n}}\partial_\chi\partial_{\chi^\dagger}
\right)\delta\zeta=\\
&
-2c_{\mathfrak{n}*}(\mathfrak{n}-1)^2(\mathfrak{n}-1)^2 l_{1\mathfrak{n}} \delta c_\mathfrak{n}^{(0)} \rho^{\mathfrak{n}-3}
\, .
\end{split}
\label{eq:zepsexplinflow}
\end{align}
We first look for a special solution of this linear inhomogeneous 
partial differential equation, which in the spherical coordinates
(\ref{eq:radius_angle}) is $\vartheta$-independent.
The radial ansatz leads to a first order ordinary differential
equation for $\delta\zeta'(\kern -0.9pt \Call{r}\, )$.
It possesses a one-parameter family of solutions,
spanned by the same additive term of \Eqref{eq:term_notto_add}.
As for the fixed point solution, we set this term to zero.
This leads to the following radial solution  
\begin{equation}
\delta\zeta'(\kern -0.9pt \Call{r}\, )=2\delta\eta\, \sum_{i=1}^{\mathfrak{n}-3}
\frac{\kern -0.9pt \Call{r}}{i!}\left(\frac{2\Call{r}^{\,2}}{\mathfrak{n}-2}\right)^{i-1}\, .
\label{eq:zepsexplinsol}
\end{equation}
Here we have already imposed that $\delta\zeta'(\kern -0.9pt \Call{r}\, )$ be smooth at the
origin, which puts a constraint on $\delta\eta$, namely
\begin{equation}
\delta\eta=c_{\mathfrak{n}*}l_{1\mathfrak{n}}l_{0\mathfrak{n}}^{\mathfrak{n}-3}
\frac{(\mathfrak{n}-1)^2(\mathfrak{n}-2)^{\mathfrak{n}-2}(\mathfrak{n}-2)!}{2^{\mathfrak{n}-4}}
\delta c_\mathfrak{n}^{(0)}\, .
\end{equation}
Compatibility between this relation and \Eqref{eq:deltac_deltaeta}
determines $\lambda=\omega$ to be 
\begin{equation}
\omega=(\mathfrak{n}-2)\epsilon,
\label{eq:omega_genericn}
\end{equation}
where we have used \Eqref{eq:FPc_epsilon}.
The fact that this result turns out to be universal suggests that it
might agree with full perturbative computations.

Let us stress that the eigensolution~\eqref{eq:zepsexplinsol}
is a polynomial in $\Call{r}$, while it shows
a branch cut at the origin if expressed in terms of $\rho$.
Since \Eqref{eq:zepsexplinsol} represents a particular solution,
one can construct the general eigensolution by adding
the general solution of the associated homogeneous equation.
The latter has, after separation of variables, the same form as \Eqref{eq:radgausslinflow},
but with $\lambda=0$ and $d=d_\mathfrak{n}$.
Since the only polynomial solutions are the ones in \Eqref{eq:gausslinsoldzeta},
for which $\lambda_{km}>0$, we conclude that 
\Eqref{eq:zepsexplinsol} describes the only acceptable eigenperturbation corresponding
to the eigenvalue of \Eqref{eq:omega_genericn}.

One might be tempted to compare the plain extrapolation of~\Eqref{eq:omega_genericn}
at $\epsilon=d_\mathfrak{n}-2=2/(\mathfrak{n}-2)$
to the exact results known in two dimensions, in the hope of good
agreement for large $\mathfrak{n}$.
The agreement is not good at all, since we obtain $\omega=2$,
while for the minimal models in two dimensions
\begin{equation}
\omega=\Delta_{\Phi\Phi^\dagger}+2-d=\frac{4}{\mathfrak{n}}\, 
\end{equation}
under the assumption that the lowest irrelevant scalar operator
is related to $\Phi\Phi^\dagger$ by the action of the four supercharges~\cite{Bobev:2015jxa}.
This is because a resummation of
the $\epsilon$ expansion is needed regardless of the numerical
value of $\epsilon$ used in the plain extrapolation.
Indeed such a disagreement had already been observed for the
purely scalar models~\cite{Itzykson:1989sx}, and can be
heuristically understood by considering that the actual expansion 
parameter is the classical dimension of the coupling $c_\mathfrak{n}$,
i.e~$(\mathfrak{n}-2)\epsilon/2$.
The latter should be equal to one in two dimensions,
and thus not small.

\section{Conclusions}
\label{sec:conclusions}

Three dimensional scale-invariant QFTs
 play a fundamental role as cornerstones in advancing and testing
our understanding of strongly interacting QFTs.
For instance, 
the Ising and the Gross-Neveu universality classes
have been extensively analyzed for decades.
Instead, comparatively few studies 
have addressed the WZ model  with
four supercharges, which in three dimensions defines
what is sometimes called the supersymmetric Ising universality class.
As an example, while the $\epsilon$
expansion about four dimensions has been computed up to
six loops  for the Ising case~\cite{Schnetz:2016fhy,Kompaniets:2017yct},
it has only recently been pushed up to three loops~\cite{Zerf:2016fti} and then four loops~\cite{Zerf:2017zqi} 
for the nonsupersymmetric generalization of the present model.
Other perturbative approaches have been adopted in the supersymmetric
case. For instance, the four-dimensional WZ model  has been studied
up to four loops~\cite{Avdeev:1982jx}, and the three-dimensional
case up to two loops with the background field method~\cite{Buchbinder:2012zd}. However, we do not know of any application of these computations to
critical models.

As in the Ising phase transition, also in the supersymmetric case
the two most interesting universal quantities
are the critical exponents $\nu$ and $\eta$.
The latter is exactly determined by supersymmetry~\cite{Aharony:1997bx}, see~\Eqref{eq:etas}.
The former is defined in terms of the non-trivial 
supersymmetry-breaking relevant perturbation, 
roughly a change in the scalar mass at constant fermion bilinear.
While supersymmetry cannot determine $\nu$, it provides an exact superscaling relation
linking it to the first correction-to-scaling exponent on the supersymmetric hypersurface~\cite{talkEmergentSusy},
which we call $\omega$, namely
\begin{equation}
\omega=2-\frac{1}{\nu}\, .
\end{equation}
The computation of this observable by means of the FRG has been the case study on which we have focused this
exploratory work.

\begin{table}[!]
	\begin{tabular}{c|c|c|c|c|c}
		& O($\epsilon^2$) & O($\epsilon^3$) &O($\epsilon^4$) & Bootstrap & This Work  \\\hline
		$\omega$ & \num{0.667} & \num{0.909} & \num{0.871(1)} & \num{0.9098(20)} & \num{0.8344}
	\end{tabular}
	\caption{The supersymmetric correction-to-scaling exponent at the $\mathcal{N}=2$
		critical WZ model  in three dimensions. See Sec.~\ref{sec:conclusions}
		for explanations.}
	\label{tab:omega}
\end{table}
Before summarizing our results, let us first review what is known from the literature.
The supersymmetric critical exponent $\omega$ has been computed at three loops in the $\epsilon$
expansion in Ref.~\cite{Zerf:2016fti}, and at four loops in Ref.~\cite{Zerf:2017zqi}  which gives the result of~\Eqref{eq:epsmore}.
The numerical values that can be extracted from this expansion at
different levels of approximation
are presented in the first
three columns of Tab.~\ref{tab:omega}.
For the two-loops approximation we give the plain extrapolation at $\epsilon=1$. For the three-loops computation 
we report the Pad\'e $[1,2]$ or $[2,1]$ resummation of Ref.~\cite{Fei:2016sgs}.
Finally the four-loops result has been used with the Pad\'e approximants
$[2,2]$ or $[3,1]$ in Ref.~\cite{Zerf:2017zqi}, obtaining $0.872$ or $0.870$
respectively, which we summarize as in the third column of Tab.~\ref{tab:omega}.
The fourth column shows the prediction of the conformal bootstrap~\cite{Bobev:2015vsa}.
The last entry presents our best estimate, obtained in Sec.~\ref{sec:critical_exponent_omega}.
We are not able to estimate the systematic errors
 related to the truncation of the theory space,
since we have not collected enough data on it. 
Yet we can select this result as the most accurate because it comes from the
less restrictive truncation, accounting for a generic field-dependent K\"ahler metric.
Furthermore, we have performed a minimal-sensitivity analysis of the regulator
dependence, observing that its minimization is in fact a maximization of $\omega$.
In Sec.~\ref{sec:momentumdependent Z section}
we have also explored the alternative direction of including the momentum
dependence of the generalized K\"ahler metric. Though in this case we were able to consider only two couplings, by changing the regulator we have obtained a maximal value $\omega=0.8317$, 
which further supports the result in Tab.~\ref{tab:omega}.

In deriving these results, we have reproduced the known nonrenormalization of the
superpotential. We have also compared
our approximations to the perturbative $\epsilon$ expansion, finding that
our flow equations not only exactly capture the one-loop contribution,
but appear to perform better than the two-loops computation already in the
simple approximation of a constant wave function renormalization, see Sec.~\ref{sec:LPA$'$ section}.
This nicely illustrates how the FRG includes resummations of subsets of higher order diagrams.
Yet, the truncations we have adopted appear to be still too poor to compete with three loops or the conformal bootstrap, since we are about 8$\%$ away from the results obtained by these methods.
Higher orders of the derivative expansion might fill this gap.
A proper numerical fixed-point analysis of the flow of~\Eqref{eq:dlessflowBLPA1} for the two-point function, i.\,e.~the first order of a vertex expansion, might also yield better results.

The present FRG analysis of the critical three-dimensional WZ model  
thus leaves room for improvement in the
determination of $\omega$, and furthermore does not address the 
computation of other properties of this
scale-invariant model that can be found in the literature, 
e.\,g.~the central charge or the sphere free energy~\cite{Bobev:2015jxa,Fei:2016sgs}.
Some of these can certainly be extracted with the RG method.
It is furthermore possible to compute data on the operator product 
expansion, both within~\cite{Sonoda:1992hd,Codello:2017hhh} or beyond perturbation theory~\cite{Sonoda:1990gp,Pagani:2016pad,Pagani:2017tdr}.
We leave such endeavors for future studies.
Still, this work provides constructive evidence in favor
of the existence of scale-invariant WZ models with four supercharges,
in the form of explicit Landau-Ginzburg descriptions that go beyond the
exact constraints imposed by the nonrenormalization
of the superpotential.
In fact, we have provided an approximation of the critical K\"ahler metric
for an infinite tower of such models in continuous dimensions, as well
as results showing that the scaling properties of these fixed points
are genuinely non-Gaussian.
This has been done in greater detail in Sec.~\ref{sec:FixedPointPotential}
for the three-dimensional $\mathcal{N}=2$ case, 
where we have determined the critical K\"ahler metric by means of local 
and global numerical methods.

In Sec.~\ref{sec:Multicrit} we have also presented a partial perturbative analysis 
of multicritical models between two and
three dimensions, with superpotential $W\propto \Phi^\mathfrak{n}$, 
employing an $\epsilon$ expansion of truncated FRG equations around the
corresponding upper critical dimensions.
Apart from constructing fixed-point solutions, we have computed the exponent $\omega$
at first order in $\epsilon$, see~\Eqref{eq:omega_genericn}.
Collecting information supporting the existence of such non-Gaussian fixed points
in continuous dimensions could seem a purely academic exercise.
Yet, there might be hope to experimentally test such phenomena through intriguing 
relations between short range statistical models in fractional dimensions 
and long range ones in integer dimensions~\cite{Defenu:2014bea}.
For the multicritical models, the present study does not address many 
interesting aspects. In particular, we are not aware of a full $\epsilon$ expansion, and nonperturbative FRG analyses in two or continuous dimensions
are missing.

Of course, the RG equations we have computed can be employed to study off-critical features
of these models, such as supersymmetry breaking, or the finite temperature 
and density phase diagram. 
Also, they can be used to search for unknown critical models that are not
revealed by a simple analysis of the superpotential, such as for instance theories with
shift symmetry or with a quadratic superpotential (e.\,g. the $\mathfrak{n}=1$ and $\mathfrak{n}=2$
cases in~\Eqref{eq:etas}).
It would also be interesting to perform an FRG analysis of models with
several superfields.
Finally, similar FRG studies might shed some light on the nature of the putative minimal four-dimensional 
$\mathcal{N}=1$ superconformal theory observed in conformal-bootstrap studies, 
see Refs.~\cite{Poland:2011ey,Bobev:2015jxa,Poland:2015mta,Xie:2016hny,Li:2017ddj}.

\section*{Acknowledgments}

We profited of several discussions with
Holger Gies and Omar Zanusso on related topics.
Luca Zambelli acknowledges support by the DFG under grants
No. GRK1523/2, and Gi 328/5-2 (Heisenberg program).
Polina Feldmann acknowledges support by the Heinrich Böll Stiftung.

\appendix

\section{Dirac Conventions}
\label{sec:appconventions}

In four-dimensional flat spacetime we adopt the signature
$(1,-1,-1,-1)$
such that $\Gamma^0$ is Hermitian and $\Gamma^i$ is anti-Hermitian.
As usual, $\bar{\psi}$ denotes the Dirac conjugate $\bar\psi=\psi^\dagger
\Gamma^0$. We set
\begin{equation}
\Gamma_5 := -i\Gamma_0\Gamma_1\Gamma_2\Gamma_3.
\end{equation}
After dimensional reduction to the three-dimensional Euclidean space 
the metric has signature $(1,1,1)$, and the Dirac conjugate becomes
\begin{equation}
\bar{\psi}\equiv\psi^\dagger.
\end{equation}
The integrals over anticommuting variables, occurring in the superfield formulations of Lagrangian densities, see e.\,g. \Eqref{eq:WessZuminoSF}, denote a Berezin integration with
\begin{equation}
\on{d}^2\!\theta\equiv\on{d}\!\theta_1\on{d}\!\theta_2, \;\on{d}^2\!\bar{\theta}\equiv\on{d}\!\bar{\theta}_2\on{d}\!\bar{\theta}_1.
\end{equation}
Note that each $\on{d}\!\theta_i$, $\on{d}\!\bar{\theta}_i$ has mass dimension 
$1/2$.

For Dirac spinors and their conjugates we use the same
Fourier transform conventions as for bosons:
\begin{equation}
f(q) = \frac{1}{\sqrt{2\pi}^d}\int \!\operatorname{d}^d\!x\; f(x)\operatorname{e}^{iqx}\, .
\end{equation}
More details on conventions and computations can be found in Ref.~\cite{PolinaFeldmann:2016bfl}.

\section{LPA$\mathbf{'}$}
\label{sec:appLPA}

Within the ansatz of \Eqref{eq:truncLPA} the
bosonic block of $\Gamma_k^{(2)}$ at constant bosonic fields 
and vanishing fermionic field is sufficient to 
obtain $\partial_tZ_0$ and $\partial_t W$. It reads
\begin{equation}
\Gamma_B^{(2)}=
	\begin{pmatrix}
	q^2Z_0 & -{W^\dagger}''' f^\dagger & 0 & -{W^\dagger}''\\
	-W''' f & q^2Z_0 & -W'' & 0 \\
	0 & - {W^\dagger}'' & -Z_0 & 0 \\
	-W'' & 0 & 0 & -Z_0
	\end{pmatrix}
	\delta(p-q).
\end{equation}
The computation of $\partial_t Z_0$ proceeds by projecting \Eqref{eq:Wetterich} 
onto zero fields $\phi,\psi$ and 
auxiliary fields $f(x)=f\delta(x)$
and subsequently evaluating  $(\partial_{f^\dagger}\partial_f)$ at $f=0$. 
The off-diagonal blocks of $\Gamma_k^{(2)}$ which mix bosons and fermions 
vanish at $\psi=0$, while the fermionic block is 
not needed since it carries no dependence on $f$ or $f^\dagger$.

\section{Kähler Potential}
\label{sec:appKaehler}

To extract the flow 
of the Kähler metric from \Eqref{eq:Wetterich} within the ansatz of \Eqref{eq:truncKaehler} 
we, once more, start out from constant fields and vanishing $\psi$.
Denoted in component fields, \Eqref{eq:truncKaehler} reads
\begin{align}
\begin{split}
\Gamma_k=&\int\!\on{d}^d\!x\left\{
Z_0\left[
\zeta\,\Big(\vert\nabla\phi\vert^2+\frac{i}{2}\bar{\psi}\slashed{\sigma}\psi-\frac{i}{2}(\partial_j\bar{\psi})\sigma^j\psi-ff^\dagger\Big)\right. \right.\\ 
&\qquad\left.\left.+\frac{1}{2}\Big(\partial_\phi \zeta\left(\psi^T\sigma^2\psi f^\dagger+i\bar{\psi}\sigma^j\psi\partial_j\phi\right)+\on{h.\!c.}\Big)\right.\right.\\
&\qquad \left.\left.-\frac{1}{2}\partial_\phi\partial_{\phi^\dagger}\zeta\psi^T\sigma^2\psi\bar{\psi}\sigma^2\psi^*\right]\right. 
\\& \left. 
-\left(W'(\phi)f-\frac{1}{2}W''(\phi)\psi^T\sigma^2\psi+\on{h.\!c.}\right)\right\}.
\end{split}
\end{align}
Its second variation at constant bosonic fields and $\psi=0$ 
consists of two diagonal blocks,
\begin{align}
\begin{split}
\Gamma_B^{(2)}&=\delta(p-q)F_B(Z_0,q^2,\zeta,f,f^\dagger)\, ,\\
\Gamma_F^{(2)}&=\delta(p-q)F_f(Z_0,q_j,\zeta,f,f^\dagger)\, ,\label{blwi}
\end{split}
\end{align}
with field-dependent $\zeta$.
For more details on these matrices see  Ref.~\cite{PolinaFeldmann:2016bfl}.
The flow of the Kähler metric is 
obtained from $(\partial_{f^\dagger}\partial_f)$ at $f=0$ and reads
\begin{align}
\begin{split}
&\left(\partial_t+\frac{\Delta}{2}\big(\chi^\dagger\partial_{\chi^\dagger}-\chi\partial_\chi\big)
+\eta\right)\zeta (\chi,\chi^\dagger)=-\int_q \frac{1}{v^3}\\
&\left\{\vphantom{\frac{i}{2}}(\partial_t-q\partial_q-\eta+1) r_1\right.\\
&\left. \Big[ h^2(M+M^\dagger)|w '''|^2-h\Big((2M^{\dagger 2}+u)w '''\partial_{\chi^\dagger}  \zeta +\operatorname{h.c.}\Big)\right.
\\&\left.+(M+M^\dagger)\Big((u-2q^2h^2)|\partial_\chi \zeta |^2+hv\partial_\chi\partial_{\chi^\dagger}  \zeta \Big)\Big] \right.\\
&\left.+(\partial_t-q\partial_q-\eta) r_2\right.\\
&\left.\Big[-hu|w '''|^2+(u-2q^2h^2)\left(M^\dagger w '''\partial_{\chi^\dagger}  \zeta +\operatorname{h.c.}\right)\right.
\\&\left.+2q^2h(2|M|^2+u)|\partial_\chi \zeta |^2-uv\partial_\chi\partial_{\chi^\dagger} \zeta\, \Big]\vphantom{\frac{i}{2}}\right\}
\label{eq:Zflow}
\end{split}
\end{align}
with the abbreviations
\begin{align}
\begin{split}
h&=\zeta  +r_2,\;\;\quad\quad\quad\;\;\;M=w ''+r_1\, ,\\
u&=|M|^2-q^2h^2,\;\;\quad\;\,\, \, v=|M|^2+q^2h^2\, ,
\end{split}
\end{align}
and the notation of \Eqref{eq:int_q}.
Setting $\zeta (\phi,\phi^\dagger)=1$ recovers the LPA$'$ result. 
We have confirmed the flow of the K\"ahler metric by deriving it also from the
projection onto the fermionic kinetic term.

\subsection{Fixed Point Couplings}
The three Tables \ref{tab:apptab1}, \ref{tab:apptab2}, and \ref{tab:apptab3} contain, 
for an exemplary set of values of the prefactor $a$ in the regulator (\ref{eq:litimrega}),
the fixed point results obtained by polynomially truncating $\zeta(\rho)$ up to order $N$,
as in \Eqref{eq:poltruncKaehler}.

\begin{table}[h!]
	\begin{tabular}{l|rrrrrr}
		$N$ & \multicolumn{1}{c}{$g_*^2$} &\multicolumn{1}{c}{$\zeta_{1*}$} & \multicolumn{1}{c}{$\zeta_{2*}$} &   \multicolumn{1}{c}{$\zeta_{3*}$}  &\multicolumn{1}{c}{$\zeta_{4*}$} &\multicolumn{1}{c}{$\zeta_{5*}$}    \\\hline
		0 & \num{1.7233} &0 &0 &0 &0 &0\\
		1 & \num{2.1749} &  \num{-3.3423} &  0  & 0 & 0 & 0 \\
		2 & \num{2.1308} &  \num{ -3.0161}& \num{8.2007} & 0 & 0 & 0  \\
		3 & \num{2.1334} &  \num{-3.0357} & \num{7.7102} & \num{-14.842} & 0 & 0\\
		4 & \num{2.1325} &  \num{-3.0290} & \num{7.8779} & \num{-9.7650} & \num{132.56} & 0\\
		5 & \num{2.1328} &  \num{-3.0313} & \num{7.8194} & \num{-11.537} & \num{86.305} & \num{-1121.5}\\
	\end{tabular}
	\caption{Fixed point couplings due to polynomial truncation of $\zeta_*(\rho)$ to order $N$; $a=\num{1.1}$. \label{tab:apptab1}}
\end{table}
\vspace{0.5em}
\begin{table}[h!]
	\begin{tabular}{l|rrrrrr}
		$N$ & \multicolumn{1}{c}{$g_*^2$} &\multicolumn{1}{c}{$\zeta_{1*}$} & \multicolumn{1}{c}{$\zeta_{2*}$} &   \multicolumn{1}{c}{$\zeta_{3*}$}  &\multicolumn{1}{c}{$\zeta_{4*}$} &\multicolumn{1}{c}{$\zeta_{5*}$}    \\\hline
		0 & \num{2.0714} &0 &0 &0 &0 &0\\
		1 & \num{2.5856} &  \num{-2.7795} &  0  & 0 & 0 & 0 \\
		2 & \num{2.5501} &  \num{-2.5877}& \num{4.5557} & 0 & 0 & 0  \\
		3 & \num{2.5479} &  \num{-2.5756} & \num{4.8418} & \num{7.6240} & 0 & 0\\
		4 & \num{2.5482} &  \num{-2.5773} & \num{4.8009} & \num{6.5346} & \num{-25.835} & 0\\
		5 & \num{2.5483} &  \num{-2.5777} & \num{4.7908} & \num{6.2655} & \num{-32.219} & \num{-139.07}\\
	\end{tabular}
	\caption{Fixed point couplings due to polynomial truncation of $\zeta_*(\rho)$ to order $N$; $a=\num{1.7}$. \label{tab:apptab2}}
\end{table}
\vspace{0.5em}
\begin{table}[h!]
	\begin{tabular}{l|rrrrrr}
		$N$ & \multicolumn{1}{c}{$g_*^2$} &\multicolumn{1}{c}{$\zeta_{1*}$} & \multicolumn{1}{c}{$\zeta_{2*}$} &   \multicolumn{1}{c}{$\zeta_{3*}$}  &\multicolumn{1}{c}{$\zeta_{4*}$} &\multicolumn{1}{c}{$\zeta_{5*}$}    \\\hline
		0 & \num{2.2929} &0 &0 &0 &0 &0\\
		1 & \num{2.8865} &  \num{-2.6793} &  0  & 0 & 0 & 0 \\
		2 & \num{2.8413} &  \num{-2.4752}& \num{4.5749} & 0 & 0 & 0  \\
		3 & \num{2.8384} &  \num{-2.4624} & \num{4.8619} & \num{7.1570} & 0 & 0\\
		4 & \num{2.8391} &  \num{-2.4656} & \num{4.7905} & \num{5.3765} & \num{-39.590} & 0\\
		5 & \num{2.8392} &  \num{-2.4658} & \num{4.7847} & \num{5.2318} & \num{-42.807} & \num{-65.562}\\
	\end{tabular}
	\caption{Fixed point couplings due to polynomial truncation of $\zeta_*(\rho)$ to order $N$; $a=\num{2.3}$. \label{tab:apptab3}}
\end{table}

\section{Momentum Dependence}
\label{sec:appMomentum}

The second variation $\Gamma_k^{(2)}(p,q)$ of ansatz~\eqref{eq:ansatzMomentum} at constant $f$ and vanishing Fermi-field is block-diagonal with
\begin{multline}
\Gamma_B^{(2)}=
\left(
\begin{array}{cc}
q^2Z_0\zeta \delta(p-q) & -\kappa W'''^\dagger(p-q) f^\dagger\\
-\kappa W'''(p-q)f & q^2Z_0\zeta \delta(p-q)\\
0&-\kappa W''^\dagger(p-q)\\
-\kappa W''(p-q)&0
\end{array}\right.\\
\cdots\left.
\begin{array}{cc}
0&-\kappa W''^\dagger(p-q)\\
-\kappa W''(p-q)&0\\
-Z_0\zeta \delta(p-q)&0\\
0&-Z_0\zeta \delta(p-q)
\end{array}\right)
\end{multline}
and
\begin{equation}
\Gamma_F^{(2)}=
\begin{pmatrix}
Z_0\zeta \sigma^jq_j\delta(p-q)&\kappa W''^\dagger(p-q)\sigma^2\\
\kappa W''(p-q)\sigma^2&Z_0\zeta \sigma^{jT}q_j\delta(p-q)
\end{pmatrix},
\end{equation}
where $\kappa=(2\pi)^{-3/2}$.

\bibliography{Master_Bib}

\end{document}